\documentclass[aps,prx,floatfix,nofootinbib,notitlepage,twocolumn]{revtex4-1}
\usepackage{booktabs}
\usepackage{lipsum}
\usepackage{graphicx}
\usepackage{amsmath,amssymb,amsthm}
\usepackage{braket}
\usepackage{mathtools}
\usepackage{hyperref}
\usepackage{color}
\usepackage{float}
\usepackage{hyperref}
\usepackage[normalem]{ulem}
\usepackage{colonequals}
\usepackage{siunitx}
\usepackage{enumitem}
\usepackage{bbm}
\usepackage[dvipsnames]{xcolor}
\usepackage{tabularx}
\usepackage{mathrsfs}

\makeatletter 
    
\renewcommand\onecolumngrid{
\do@columngrid{one}{\@ne}%
\def\set@footnotewidth{\onecolumngrid}
\def\footnoterule{\kern-6pt\hrule width 1.5in\kern6pt}%
}

\renewcommand\twocolumngrid{
        \def\footnoterule{
        \dimen@\skip\footins\divide\dimen@\thr@@
        \kern-\dimen@\hrule width.5in\kern\dimen@}
        \do@columngrid{mlt}{\tw@}
}%

\makeatother    

\newtheorem{lemma}{Lemma}
\newtheorem{prop}{Proposition}

\newcommand{\ov}{\overline}

\newcommand{\C}{\mathbb{C}}
\newcommand{\F}{\mathcal{F}}
\newcommand{\LLL}{\text{LLL}}

\newcommand{\n}[1]{\left| #1 \right|}

\newcommand{\sign}{\operatorname{sign}}

\newcommand{\sz}{\mathsf{z}}
\newcommand{\spar}{\overline{\partial_{\sz}}}
\newcommand{\sk}{\mathsf{k}}

\newcommand{\su}{\mathsf{u}}
\newcommand{\sU}{\mathsf{U}}
\newcommand{\sG}{\mathsf{G}}

\renewcommand{\v}[1]{\boldsymbol{#1}}
\DeclareMathOperator{\Tr}{Tr}
\DeclareMathOperator{\tr}{tr}
\DeclareMathOperator{\diag}{diag}
\let\Re\relax
\let\Im\relax
\DeclareMathOperator{\Re}{Re}
\DeclareMathOperator{\Im}{Im}

\newcommand{\R}{\mathbb{R}}

\renewcommand{\P}{\mathcal{P}}
\newcommand{\Q}{\mathcal{Q}}
\newcommand{\T}{\mathcal{T}}
\newcommand{\Pmany}{\mathscr{P}}

\newcommand{\bz}{\mathrm{BZ}}
\newcommand{\uc}{\mathrm{uc}}
\newcommand{\FS}{\mathrm{FS}}

\newcommand{\SRGS}{SRI-GS}

\newcommand{\br}{{\v r}}

\newcommand{\bk}{{\v k}}
\renewcommand{\k}{\bk}
\newcommand{\bq}{{\v q}}

\newcommand{\ba}{{\v a}}

\newcommand{\bG}{{\v G}}

\newcommand{\bA}{{\v A}}
\newcommand{\bQ}{{\v Q}}
\newcommand{\bsigma}{{\v \sigma}}
\newcommand{\bnabla}{{\v \nabla}}

\DeclarePairedDelimiter\abs{\lvert}{\rvert}%
\DeclarePairedDelimiter\norm{\lVert}{\rVert}%
\makeatother

\begin{document}

\title{Vortexability: A Unifying Criterion for Ideal Fractional Chern Insulators}

\author{Patrick J. Ledwith}
\affiliation{Department of Physics, Harvard University, Cambridge, MA 02138, USA}
\author{Ashvin Vishwanath}
\affiliation{Department of Physics, Harvard University, Cambridge, MA 02138, USA}
\author{Daniel E. Parker}
\affiliation{Department of Physics, Harvard University, Cambridge, MA 02138, USA}

\begin{abstract}
Fractional Chern insulators realize the remarkable physics of the fractional quantum Hall effect (FQHE) in crystalline systems with Chern bands. The lowest Landau level (LLL) is known to host the FQHE, but not all Chern bands are suitable for realizing fractional Chern insulators (FCI). Previous approaches to stabilizing FCIs focused on mimicking the LLL through momentum space criteria. Here instead we take a real-space perspective by introducing the notion of vortexability. Vortexable Chern bands admit a fixed operator that introduces vortices into any band wavefunction while keeping the state entirely within the same band. Vortexable bands admit trial wavefunctions for FCI states, akin to Laughlin states. In the absence of dispersion and for sufficiently short ranged interactions, these FCI states are the ground state  --- independent of the distribution of Berry curvature. Vortexable bands are much more general than the LLL, and we showcase a recipe for constructing them. We exhibit diverse examples in graphene-based systems with or without magnetic field, and with any Chern number. A special class of vortexable bands is shown to be equivalent to the momentum space ``trace condition" or ``ideal band condition". In addition, we also identify a more general form of vortexability that goes beyond this criterion. We introduce a modified measure that quantifies deviations from general vortexability which can be applied to generic Chern bands to identify promising FCI platforms.

\end{abstract}

\maketitle

\section{Introduction}

  Widely considered to be the most remarkable example of emergence, the fractional quantum Hall effect (FQHE) has traditionally been studied at partial fillings of the lowest Landau level (LLL). A Chern band \textit{may} also host a gapped lattice-translation symmetric FQHE state at partial filling; these are translation symmetry-enriched topological orders\cite{parkerFieldtunedZerofieldFractional2021,spantonObservationFractionalChern2018,thorngrenGaugingSpatialSymmetries2018,elseNonFermiLiquidsErsatz2021,macdonaldLandaulevelSubbandStructure1983,luSymmetryprotectedFractionalChern2012}, known as Fractional Chern Insulators (FCIs) \cite{parameswaranFractionalQuantumHall2013a, bergholtzTOPOLOGICALFLATBAND2013,liuRecentDevelopmentsFractional2022,neupertFractionalQuantumHall2011,shengFractionalQuantumHall2011,regnaultFractionalChernInsulator2011,scaffidiAdiabaticContinuationFractional2012a,grushinEnhancingStabilityFractional2012,kourtisFractionalChernInsulators2014}. FCIs can form at any magnetic field --- including zero. Their spectral gap $E_{\mathrm{FCI}} \sim 1/a$ is set by the unit cell scale, rather than $E_{\mathrm{LLL}} \sim \sqrt{B}$ in the FQHE, bringing the promise of enhanced energy scales $E_{\mathrm{FCI}}\gg E_{\mathrm{LLL}}$. FCIs were experimentally reported \cite{spantonObservationFractionalChern2018} in Harper-Hofstadter bands \cite{scaffidiExactSolutionsFractional2014a,mollerFractionalChernInsulators2015,andrewsStabilityFractionalChern2018,andrewsFractionalQuantumHall2020a,andrewsStabilityPhaseTransitions2021,bauerFractionalChernInsulators2022}. They were also recently predicted to appear in the Chern bands of twisted bilayer graphene \cite{ledwithFractionalChernInsulator2020a, abouelkomsanParticleHoleDualityEmergent2020,repellinChernBandsTwisted2020,wilhelmInterplayFractionalChern2021a}, and subsequently observed with the help of a small magnetic field \cite{xieFractionalChernInsulators2021,parkerFieldtunedZerofieldFractional2021}. 
  
  While the energetics of the FQHE in the LLL are well established, the energetics of FCIs are much less understood due to the enormous diversity of Chern bands and the generic lack of any analytic control over the single particle wavefunctions. How do we find Chern bands that are likely to host FCI states?

One popular --- but ultimately limiting --- approach to finding candidate Chern bands for FCIs is to completely mimic the LLL through its momentum space band geometry \cite{jacksonGeometricStabilityTopological2015, claassenPositionMomentumDualityFractional2015, leeBandStructureEngineering2017,ledwithFractionalChernInsulator2020a,meraEngineeringGeometricallyFlat2021, meraKahlerGeometryChern2021, ozawaRelationsTopologyQuantum2021, meraRelatingTopologyDirac2021, zhangRevealingChernNumber2021,varjasTopologicalLatticeModels2021a}. The momentum space band geometry (MSBG) of the LLL has a host of special properties: uniform Berry curvature, the ``determinant" and ``trace" conditions, the GMP algebra  \cite{girvinMagnetorotonTheoryCollective1986}, and more.  While it is possible to completely reproduce LLL physics by satisfying these conditions \cite{royBandGeometryFractional2014a}, the principle of mimicry is insufficient to determine which conditions are central versus merely peripheral.

Hints at a deeper perspective come from recent works on chiral twisted graphene \cite{tarnopolskyOriginMagicAngles2019a,ledwithFractionalChernInsulator2020a,popovHiddenWaveFunction2021,shefferChiralMagicAngleTwisted2021a,renWKBEstimateBilayer2021a,wangChiralApproximationTwisted2021a,naumisReductionTwistedBilayer2021a,navarro-labastidaWhyFirstMagicangle2022,beckerSpectralCharacterizationMagic2021a,beckerFineStructureFlat2022,beckerIntegrabilityChiralModel2022}, which focused on the ``trace condition" \cite{ledwithFractionalChernInsulator2020a,ledwithFamilyIdealChern2022,wangExactLandauLevel2021a,ledwithStrongCouplingTheory2021}. The exactly flat Chern bands of chiral twisted graphene not only satisfy the trace condition but also have a transparent \emph{real-space} structure. This structure enables the construction of ``short range ground states (\SRGS{})": 
FQHE trial wavefunctions that are exact many-body ground states of short-range (pseudopotential) interactions \cite{trugmanExactResultsFractional1985a,ledwithFractionalChernInsulator2020a,ledwithStrongCouplingTheory2021,wangExactLandauLevel2021a,dongDiracElectronPeriodic2022}. Subsequent work have shown the trace condition directly enables \SRGS{} construction \cite{ledwithFamilyIdealChern2022}. Is the trace condition essential, or one example of a deeper structure?

We will adopt a unifying real-space perspective. After all, the rich geometrical structures of the FQHE are rooted in real space, as are most trial wavefunctions. We will show how to place the real-space geometry of FCIs on the same footing as the FQHE.
Earlier work has developed different aspects of the geometry of the fractional quantum Hall effect in the LLL
\cite{haldaneGeometricalDescriptionFractional2011,canFractionalQuantumHall2014,canGeometryQuantumHall2015,klevtsovGeometricAdiabaticTransport2015,douglasBergmanKernelPath2009,gromovBimetricTheoryFractional2017,gromovInvestigatingAnisotropicQuantum2017,abanovElectromagneticGravitationalResponses2014,huMicroscopicDiagnosisUniversal2021,lapaGeometricQuenchFractional2019,gromovDensityCurvatureResponseGravitational2014,gromovBoundaryEffectiveAction2016,youTheoryNematicFractional2014,bradlynTopologicalCentralCharge2015,gromovFramingAnomalyEffective2015,maciejkoFieldTheoryQuantum2013,schineSyntheticLandauLevels2016,schineElectromagneticGravitationalResponses2019}.
 Here, we describe an alternate approach to real-space geometry of FCIs that allows for continuous, discrete, or even absent
translation symmetry, and captures sub unit-cell length
scales that are known to be crucial for FCI energetics \cite{trugmanExactResultsFractional1985a}. In contrast to much of the FCI
literature \cite{parameswaranFractionalQuantumHall2013a,bergholtzTOPOLOGICALFLATBAND2013,liuRecentDevelopmentsFractional2022}, we do not work directly in the singular tight-binding limit; 
instead we use continuum models, the setting in which the energetics and geometry of the FQHE is most transparent.

Let us recall the construction of the Laughlin state in the LLL,  a \SRGS{}, on the infinite planar (or expandable disk) geometry. We begin with the fully filled state $\ket{\Psi_{\nu=1}}$ and attach a Jastrow factor that makes each particle see every other particle as an order $2s$ vortex:
\begin{equation}
  \ket{\Psi^{(2s)}} = \prod_{i<j} (z_i - z_j)^{2s} \ket{\Psi_{\nu=1}}.
    \label{eq:Laughlin}
\end{equation}
Here $z = x+iy$ and $s$ is a positive integer.
The construction \eqref{eq:Laughlin} produces the Laughlin state at filling $\nu = 1/m$ with $m=2s+1$. Note that the particle density has changed as a result of the Jastrow prefactor $\prod_{i<j} (z_i - z_j)^{2s}$ which pushes particles apart --- crucial for minimizing the Coulomb repulsion. This intuition may be made precise: the Laughlin state is a \SRGS{} \cite{haldaneFractionalQuantizationHall1983a,trugmanExactResultsFractional1985a}.

It is tantalizing to apply this construction to any Chern band. However, Eq. \eqref{eq:Laughlin} required a crucial single particle property of the LLL: attaching vortices (i.e., multiplying by $z$) keeps the wavefunction within the LLL (Fig. 1; details to follow) --- a prerequisite to construct FQHE trial states. By contrast, attaching vortices to generic Chern bands takes the wavefunction outside the band of interest, thereby greatly increasing kinetic energy. Our goal is to emphasize the property of vortex attachment and generalize it beyond the lowest Landau level.

 In this paper we introduce \textit{vortexable bands}, a class of bands to which one can attach vortices while remaining within the subspace defined by the bands. Remarkably, vortexable bands are much more general than the lowest Landau level. Below we provide examples both with or without a magnetic field; with continuous, discrete, or absent translation symmetry;
 and for any Chern number. In such vortexable bands, we can construct generalized FQHE trial states which include exact many-body \SRGS{}, like Laughlin states in the lowest Landau level.

Real space vortexability provides an organizing principle for many ideas in momentum space band geometry. We show that a subset of \textit{special} vortexable bands, where the ``vortex function" is linear, are exactly the trace condition bands. However, \textit{general} vortexable bands, with nonlinear vortex functions are beyond the scope of traditional band geometry --- but appear often. For example, we find spatially varying strain in graphene yields a general vortexable band. This shortcoming of band geometry is rooted in the implicit \textit{choice} to define the periodic part of the Bloch wavefunction: $u_{\v{k}}(\v{r}) = e^{-i\v{k}\cdot\v{r}} \psi_{\v{k}}(\v{r})$ using a plane wave $e^{-i \bk \cdot \br}$. In fact, any function $e^{-i \bk \cdot \v{\phi}(\br)}$ which satisfies  $ \v{\phi}(\br+\ba) = \v{\phi}(\br) + \ba$ under lattice translations will suffice. General vortexable bands leverage precisely this freedom to expand the space of vortexable bands. We conclude with a momentum space condition to detect all vortexable bands.

\section{Single-Particle structure of vortexable bands}

\begin{figure}
    \centering
    \includegraphics[scale=1.2]{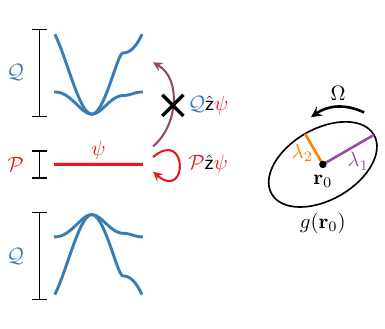}
    \caption{(Left) A vortexable band \eqref{eq:zpsiPzpsi} with associated projector $\mathcal{P}$ and complement $\mathcal{Q} = I - \mathcal{P}$. Acting with $\hat{\sz}$ keeps the wavefunction within the vortexable band: $\hat{\sz} \psi = (\mathcal{P}  + \mathcal{Q}) \hat{\sz} \psi  = \mathcal{P} \hat{\sz} \psi$.
    (Right) The vortex metric $g(\v{r}_0)$ from Eq. \eqref{eq:vortex_geometry} describes the shape of the dip in probability density induced by the vortex attachment \eqref{eq:vortex_at_r0},\eqref{eq:vortex_shape}.}
    \label{fig:ProjectorCartoon}
\end{figure}

	Consider a Chern $C$ band\footnote{We use the term ``band" for linguistic convenience; by ``band" we mean any extensive subspace separated by spectral gaps below and above so that the Hall conductance changes by $Ce^2/h$ upon fully filling the band, giving a Chern number $C$. Thus our statements and techniques apply equally well to single or multiple Bloch bands with any number of non-positional orbitals (e.g. layers), or even bands without any translation symmetry.} with an orthonormal basis $\{ \psi_{\alpha}(\v{r}) \}$ of (single-particle) continuum wavefunctions on $\R^2$. Let $\mathcal{P} = \sum_{\alpha} \ket{\psi_\alpha} \bra{\psi_\alpha}$ be the projector onto the band of interest (Fig \ref{fig:ProjectorCartoon}).

A band is \textit{vortexable} with vortex function $\sz: \R^2 \to \C$ if 
\begin{equation}
  \sz(\br) \ket{\psi} = \P \sz(\br) \ket{\psi},
 \label{eq:zpsiPzpsi}
\end{equation}
for any $\ket{\psi}$ in the band, i.e. $\mathcal{P} \ket{\psi} = \ket{\psi}$. Vortex functions $\sz(\v{r})$ can be thought of as complex coordinates on the plane: they act like $x+iy$, but need not be equal to $x+iy$ (technical definition below). A simple class of possible vortex functions is $\sz(\v{r}) = x+iy + \delta(\br)$, where $\delta(\br)$ is a sufficiently-small deformation of the standard vortex $x+iy$. Note that $\sz(\br)$ may be nonlinear in $\br$. 

Multiplying a vortexable band by the operator $\sz(\v{r})$ keeps the wavefunction entirely within the band (Fig. \ref{fig:ProjectorCartoon}) --- a strong condition few bands satisfy. 
The first example of a vortexable band is the LLL itself. Recall that vortex insertion in the LLL proceeds by adiabatically inserting a $2\pi$ flux, which has the effect of shifting angular momentum of each of the single particle states $\psi_m \propto z^m e^{-z\bar{z}/4}$ by $m \to m+1$. Clearly this can also be achieved by multiplying the wavefunction by $z$. Individual higher Landau levels are not vortexable, but the first $C$ Landau levels together comprise a Chern $C$ vortexable band. Our construction Eq. \eqref{eq:zpsiPzpsi} generalizes this idea well beyond the LLL. We now describe the real-space geometry of vortexable bands and some technical properties; eager readers may skip to the construction of SRGS trial wavefunctions in the next section.

Historically, Eq. \eqref{eq:zpsiPzpsi} first appeared with $\sz = x+iy$ as an intermediate step to reproduce the GMP algebra\cite{girvinMagnetorotonTheoryCollective1986} of the LLL from MSBG in translation symmetric bands \cite{royBandGeometryFractional2014a}. However, reproducing the GMP algebra relies on both uniform Berry curvature and \eqref{eq:zpsiPzpsi}. The condition \eqref{eq:zpsiPzpsi} reappeared in Ref. \cite{ledwithFamilyIdealChern2022} by some of us, where its relationship to band geometry independent of Berry curvature distribution was highlighted, chiral MATBG was shown to satisfy it, and its role in ensuring FQH trial states are also exact ground states with short ranged interactions was mentioned.

\subsection{Vortex Geometry}

  Consider attaching a vortex at a generic point $\br_0$:
  \begin{equation}
    \psi \mapsto \tilde{\psi} = (\sz(\br) - \sz(\br_0)) \psi.
    \label{eq:vortex_at_r0}
  \end{equation}
  We call this ``vortex attachment" because $\tilde{\psi}$ has a zero at $\br_0$ and winds by a $\pm 2 \pi$ phase around it. This creates a dip in probability density near the position $\br_0$ where the wavefunction vanishes; at the many body level these will become ``correlation holes". Near the point $\br_0$ we have
    \begin{equation}
      \abs{\tilde{\psi}(\br-\br_0)}^2 = g_{\mu\nu}(\br_0) r^\mu r^\nu \, \abs{\psi(\br_0)}^2 + O(\br^3),
      \label{eq:vortex_shape}
    \end{equation}
where $g_{\mu \nu}(\br) = \Re \partial_\mu \ov{\sz}(\br) \partial_\nu \sz(\br)$ is a ``vortex metric" that describes the elliptical shape of the vortex core, such that the semimajor and semiminor lengths $\lambda_{1,2}(\v{r}_0)$ are the eigenvalues of $g(\v{r}_0)$. We note the overall scale of $\sz$ and $g$ is not fixed.

    To eliminate pathological cases, we impose a  non-degeneracy condition. A function $\sz: \mathbb{R}^2 \to \mathbb{C}$ is a \textit{vortex function} if the vortex-metric eigenvalues are uniformly bounded above and below: 
\begin{equation}
	c^{-1} \le \lambda_{1,2}(\v{r}) \le c.
	\label{eq:non-degeneracy_condition}
\end{equation}
This ensures the vortex cores do not get arbitrarily large, small, or squeezed. 
In fact, \eqref{eq:non-degeneracy_condition} implies
  the vortex function $\sz: \mathbb{R}^2 \to \mathbb{C}$ is invertible, such that there exists coordinates $\br'(\br) = (x', y')$ where $\sz = x' + i y'$, through a global version of the inverse function theorem, see Ref.~\onlinecite{demarcoGlobalInversionFunctions1994}. 
    
    We caution that the diffeomorphism  $\v{r} \mapsto {\v{r}}'$ does \textit{not} mean that any vortexable band is physically equivalent to one with $\sz = x + iy$. This is because distances are not preserved under the diffeomorphism so, e.g., Coulomb interactions $V(r) = 1/\norm{\br} \neq 1/\norm{\br'}$ single out the ``laboratory" frame, which we use unless otherwise stated. However, the $\br'$ coordinates will give a convenient link to momentum space band geometry.

  The vortex metric is part of a Hermitian metric:
  \begin{equation}
    \begin{aligned}
      \eta_{\mu \nu}(\br) &  = \partial_\mu \ov{\sz(\br)} \partial_\nu \sz(\br), \\g_{\mu \nu}(\br)  = \Re \eta_{\mu \nu}(\br),
     & \quad \Omega(\br) \varepsilon_{\mu \nu}  =  \Im \eta_{\mu \nu}(\br).
    \end{aligned}
    \label{eq:vortex_geometry}
  \end{equation}
  We call $\Omega$ the \emph{vortex chirality}
because the phase winding of vortices is $\sign(\Omega) 2\pi$, which reverses under $\sz \leftrightarrow \overline{\sz}$.
    The non-singular metric $g$ implies that $\Omega$ does not vanish or change sign, such that the vortex phase winding is the same everywhere, because the rank one nature of $\eta$ implies $\det g = \abs{\Omega}^2$.
    In fact, vortex chirality is tied to Chern number: $\Omega C \geq 0$: we show this in the first section of the supplementary material (SI).
    We note in passing that the condition $\det g = \abs{\Omega}^2$  specifies a real-space K\"ahler structure, reminiscent of that of MSBG \cite{meraKahlerGeometryChern2021}.

\subsection{Uniqueness}
\label{subsec:uniqueness}
Can a vortexable band have two or more distinct vortex functions? With relatively weak assumptions we can show that this is not the case. In the SI we show that the vortex function is unique up to affine transformations $\sz \to \alpha \sz + \beta$ if the following conditions hold (Prop \ref{prop:uniqueness}:
\begin{enumerate}
    \item The vortexable band has a discrete translation symmetry with translation operator $\T_\ba$, in the sense that $[\T_\ba , \P] = 0$ and $\partial_\mu \sz(\br+\ba) = \partial_\mu \sz(\br)$
    \item The electron density at full filling, $\rho(\v{r}) =\sum_\alpha \n{\psi_\alpha(\br)}^2$, is finite, continuous, and almost always nonzero.
\end{enumerate}

Let us briefly comment on the two conditions. We view the discrete translation symmetry condition as a technical tool but not a strong physical restriction. Indeed, we make no assumption on how large the unit cell is or how many Bloch bands are encompassed by the vortexable ``band", and we expect most condensed matter systems to be translation symmetric in the limit of thermodynamically large unit cells. A condition on the density $\rho(\br)$ is necessary to eliminate pathological density profiles: it is always possible to modify the vortex function on regions of vanishing electron density.

    \section{Vortexable bands admit pseudopotential FQH Trial States}
We now construct FQHE ground states from vortexable bands, and demonstrate they are exact many-body ground states for short range (pseudopotential) interactions (\SRGS{}).
       At the single particle level, iterating \eqref{eq:zpsiPzpsi} implies
       \footnote{Note that $f(\sz)$ is not a ``new'' vortex function, since $f$ must be linear else the non-degeneracy condition is not satisfied. Indeed, $f'(\sz)$ is either constant or unbounded by Liouville's theorem, and the metric associated to $f(\sz)$ is $\abs{f'(\sz)}^2 \partial_\mu \ov{\sz} \partial_\nu \sz + \text{c.c.}$.}
    \begin{equation}
      f(\sz) \psi = \P f(\sz) \psi \text{ for holomorphic } f(\sz) = \sum_{n \geq 0} f_{n} \sz^n.
      \label{eq:any_fofz}
    \end{equation}
	Next, at the many-body level, we may apply \eqref{eq:any_fofz} to each particle of the system to obtain
    \begin{equation}
      f(\sz_1, \ldots, \sz_N) \ket{\Psi} = \Pmany f(\sz_1, \ldots, \sz_N) \ket{\Psi},
      \label{eq:any_fofz_manybody}
    \end{equation}
    where $\Pmany = \otimes_{i} \P_i$ is the many body projector, $\sz_i = \sz(\br_i)$ is the vortex operator for particle $i$, $f$ is holomorphic in each $\sz_i$, and $\ket{\Psi} = \Pmany \ket{\Psi}$ is any initial state of electrons in the band of interest. The ability to attach holomorphic functions to many body wavefunctions in vortexable bands generalizes any trial wavefunction in the LLL to any vortexable band. We focus on Laughlin-like states
    \begin{equation}
      \ket{\Psi^{(2s)}_{\sz}} = \prod_{i<j} (\sz_i - \sz_j)^{2s} \ket{\Psi},
    \label{eq:generalized_Laughlin}
    \end{equation}
    so we may make pseudopotential arguments, but we imagine future works will find interest in examining composite Fermi liquids or Pfaffian states in vortexable bands. We additionally emphasize that while we are inspired by the Laughlin construction, the construction \eqref{eq:generalized_Laughlin} applies to any vortexable band, and can lead to states qualitatively different than Laughlin states (for example, in vortexable bands with higher Chern number).

    As two particles $i$ and $j$ approach each other at a center-of-mass position $\br_0$, the probability density associated to \eqref{eq:generalized_Laughlin} vanishes as 
    \begin{equation}
      \abs{\Psi^{(2s)}_{\sz}}^2 \propto \abs{g_{\mu \nu}(\br_0) r_-^\mu r_-^\nu}^{4s},
      \label{eq:wf_vanish_as}
    \end{equation}
	where $\br_- = \br_i - \br_j$ is the relative coordinate. The high power $4s$ by which \eqref{eq:generalized_Laughlin} vanishes ensures that $\ket{\Psi^{(2s)}_{\sz}}$ is a SRGS\cite{trugmanExactResultsFractional1985a}; see the supplementary material for a pedagogical review of this argument. Note that we focus on a density-drensity interaction Hamiltonian --- the vortexable band should have zero bare dispersion. As we will see in the examples section, this is typically automatic.

    The reader may worry that vortex attachment would also generate FCI ground states in a topologically trivial band ($C=0$). In the first section of the SI we show that $C=0$ vortexable bands instead admit a basis of delta-function-localized position eigenstates (similar to a single band tight binding model or a free particle). The ability to ultra-localize electrons in a band projected manner implies that it is straightforward to keep them apart. Indeed, for $C=0$ vortexable bands a simple charge density wave is a zero-energy ground state of any potential with range strictly less than the interparticle spacing.

    We have argued that topological vortexable bands with zero dispersion host FCI ground states in the limit of short range interactions. However, this limit may be hard to reach if the Berry curvature is highly inhomogeneous. Indeed, consider the extreme limit of a soleinoidal Berry curvature. For a finite size system with periodic boundary conditions, such that momentum space is discretized, the Berry curvature is completely invisible to the momentum space grid. In this case the band can energetically resemble a trivial band, see e.g. Ref. \cite{khalafSoftModesMagic2020a}.

  \section{Chiral models as an origin of vortexability}

  We have shown that vortexable bands generalize the LLL in such a way to preserve vortex attachment and the ensuing energetic arguments for hosting \SRGS{}. However, vortexable bands can and do arise in systems quite dissimilar to free electrons in a magnetic field. We now consider four examples of vortexable bands: two have $\sz \neq x+iy$, two have no magnetic field, and one has arbitrary Chern number $C > 1$.

  Our examples have a common origin: a continuum chirally symmetric model $\{H, \sigma_z \}  = 0$. These models have the general form
   \begin{equation}
    H = \begin{pmatrix} 0 & D^\dag \\ D & 0 \end{pmatrix}, \qquad 
    \sigma_z =  \begin{pmatrix} I & 0 \\ 0 & -I \end{pmatrix},
    \label{eq:chiral_ham_general}
  \end{equation}
  for some operator $D$. If $D$ satisfies both of the following two conditions then the space of zero modes of $D$ is a vortexable band.
  \begin{enumerate}
    \item[P1.] $[D, \sz(\br)] = 0$ for a vortex function $\sz$.
    \item[P2.] $D$ has an extensive number of zero modes.
  \end{enumerate}
  Property P1 holds when the dependence of $D$ on derivatives $\partial_x$ and $\partial_y$ is through a single differential operator with complex valued-coefficients $\overline{\partial_{\sz}}$. Solving the Beltrami equation $\overline{\partial_{\sz}} \sz = 0$ then yields a vortex function that satisfies $[D, \sz] =0$.

  \subsection*{E1: Strained Graphene in Magnetic Field (Nonlinear \texorpdfstring{$\sz(\v{r})$}{z(r)}).}
  Our first example demonstrates that nonlinear vortex functions $\sz(\br)$ arise naturally in strained graphene.
  When placed on a bumpy surface, or subject to external stress, the atoms of a graphene lattice are displaced by $\v{u}(\v{r})$ which modifies the lab-frame metric of the graphene surface. Indeed, the graphene sheet is an embedded surface with (lab frame) metric $g_{\mu\nu}(\v{r}) = \delta_{\mu\nu} + 2 u_{\mu\nu}(\v{r}) = \delta_{ab} e_{\mu}^a(\br) e_{\nu}^b(\br)$ where $u_{\mu\nu} = [\partial_\mu u_{\nu} + \partial_{\nu} u_{\mu}]/2$ is the strain tensor and $e^a_{\mu}(\v{r}) = \delta^a_{ \mu} + \delta^{a \nu} u_{\mu\nu}$. Here $\delta$ is always the Kronecker delta regardless of index positions. We will use shortly use the inverse metric $g^{\mu \nu}$ to raise indices: $e^\mu_a = g^{\mu \nu} e_\nu^a$. The latin index positions do not matter (they are raised and lowered by $\delta^{ab}$). 
  
  The uneven distances between atomic sites modifies the velocity operator to $\hat{v}^\mu = v e^\mu_a(\br) \sigma^a$, whose anisotropy and position dependence is encoded into the orthonormal basis $e^\mu_a(\br)$.
  The continuum Hamiltonian is then
\cite{suzuura2002phonons,manesSymmetrybasedApproachElectronphonon2007,kimGrapheneElectronicMembrane2008,guineaGaugeFieldInduced2008,pereiraStrainEngineeringGraphene2009,vozmedianoGaugeFieldsGraphene2010,de2012space,manesGeneralizedEffectiveHamiltonian2013,de2013gauge,namLatticeRelaxationEnergy2017,koshinoEffectiveContinuumModel2020}.
  \begin{equation}
    H_{\text{strained graphene}}(\v{r}) = v e^\mu_a(\br) \sigma^a(-i\partial_\mu - A^{\rm{eff}}_\mu),
    \label{eq:curved_space_graphene}
\end{equation}
where $A^{\rm{eff}}_\mu = \mathcal{A}^{\rm el}_\mu(\v{r}) + A^{\rm ext}_\mu(\v{r})$. Here $A^{\rm ext}_\mu(\v{r})$ is the magnetic vector potential and $\mathcal{A}^{\rm el}_\mu(\v{r})$ is a pseudomagnetic vector potential due to the elastic modulation of the hopping matrix elements $t(a_0) \mapsto t(a_0 + u(\v{r}))$. Note that there are also other terms that are second order in gradients that arise from the hopping modulation \cite{de2012space,de2013gauge,VafekKangContinuumEffective2022,kangPseudomagneticFieldsParticlehole2022}; we ignore them here for simplicity though they do not change our conclusions which will be based on topology and the Dirac nature of \eqref{eq:curved_space_graphene}.

When written in the form \eqref{eq:chiral_ham_general}, the zero  mode operator $D$ of \eqref{eq:curved_space_graphene} is 
\begin{equation}
    D_{\text{strained graphene}} = 2v\chi^\mu(\br)( -i \partial_\mu - A_\mu^{\rm{eff}}),
\end{equation}
where $\chi^\mu = \frac{1}{2}(e^\mu_1 + i e_2^\mu)$. 
The vortex operator is now obtained by solving the Beltrami equation 
\begin{equation}
    \chi^\mu(\br) \partial_\mu \sz = \spar \sz = 0,
\end{equation}
such that property P1 is satisfied.

The easiest way to achieve P2 is a magnetic field $B_0 \hat{z} = \nabla \times A^{\rm{ext}}$, but experiments have also found strain Landau levels with $\mathcal{A}^{\rm{el}} \neq 0$, $A^{\rm{ext}} = 0$ \cite{levy2010strain}. In either case, the number of zero modes is tied to the number of flux quanta 
by an index theorem \cite{shefferChiralMagicAngleTwisted2021a} (see e.g. \cite{nakaharaGeometryTopologyPhysics2003} Eq. 12.86).
\begin{equation}
  \text{dim} \ker D - \text{dim} \ker D^\dag = N_\phi = \int \bnabla \times \bA^{\rm eff} d^2 \br,
  \label{eq:index}
\end{equation}
such that the number of zero modes $\text{dim}\ker D$ is bounded below by $N_\phi$, which is typically extensive in the system size. Therefore Eq. \eqref{eq:curved_space_graphene} gives an experimentally-realizable vortexable band without any translation symmetry and vortex $\sz \neq x+iy$. Indeed, $\sz(\v{r})$ is typically nonlinear in $\v{r}$.

  \begin{figure}
      \centering
      \includegraphics[width=\linewidth]{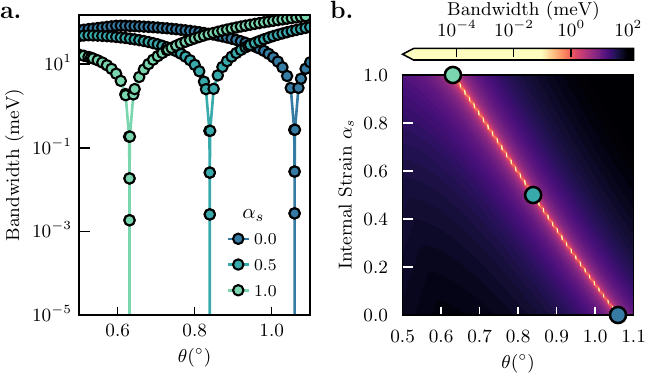}
      \caption{The ``magic line" of chiral twisted graphene. Bandwidth of chiral TBG, Eq. \eqref{eq:strained_chiral_limit}, as a \textbf{a}. discrete or  \textbf{b}. continuous function of internal strain and angle. We take a realistic \cite{koshinoEffectiveContinuumModel2020} internal strain vector potential and modulate its strength as $\alpha_s \mathcal{A}(\v{r})$, where $\alpha_s$ interpolates from $0$ to the realistic value of $1$. Details in the text. 
      }
      \label{fig:internal_strain_vortexable_line}
  \end{figure}
  
  \subsection*{E2: Chiral Twisted Bilayer Graphene (no magnetic field)}

  We will now show how chiral twisted bilayer graphene yields a vortexable band with no magnetic field. Ordinary twisted bilayer graphene (TBG) contains tunneling matrix elements for inter-sublattice tunneling and intra-sublattice tunneling. 
However the energetic preference for local AB over AA stacking induces lattice deformations
\cite{carrMinimalModelLowenergy2019,koshinoEffectiveContinuumModel2020,canteleStructuralRelaxationLowenergy2020a,lucignanoCrucialRoleAtomic2019a}, leading to a suppression of the AA tunneling, as well as an experimentally-unavoidable internal strain field that we will discuss later. 
The suppression of the AA tunneling motivated Ref. \cite{tarnopolskyOriginMagicAngles2019a} to consider an idealized limit of \emph{zero} AA tunneling. The Hamiltonian is then of the form \eqref{eq:chiral_ham_general} with
  \begin{equation}
  D_{\rm{TBG}} = 
      \begin{bmatrix}
       -2iv \ov{\partial} & w_1 U_1(\v{r})\\
      w_1 U_1(-\v{r}) & -2iv \ov{\partial}.
      \end{bmatrix},
      \label{eq:unstrained_chiral_limit}
  \end{equation}
  where    $U_1(\v{r}) = \sum_{n=0}^2 e^{i n \phi-i\v{q}_n \cdot \v{r}}$, $\v{q}_n = C_{3z}^{n} k_{\theta} (0,-1)$, $\phi = \tfrac{2\pi}{3}$, $k_\theta = \vert\v{K}^{\rm{top}} - \v{K}^{\rm{bot}}\vert$, and $w_1 \approx \SI{110}{meV}$ is the AB interlayer tunnelling. Note that $D_{\text{TBG}}$ only depends on the antiholomorphic derivative $\ov{\partial} = \frac{1}{2}(\partial_x + i \partial_y)$.
  The spectrum of the Hamiltonian defined with \eqref{eq:unstrained_chiral_limit} depends only on the dimensionless parameter $\alpha = w_1/v_F k_{\theta}$, up to an overall scale $v_F k_\theta$. 
  
  The operator \eqref{eq:unstrained_chiral_limit} satisfies P1 for $z = x+iy$ and, at the `magic angle' $\alpha \approx 0.586$, P2 is satisfied as well. Here $D_{\mathrm{TBG}}$ has an extensive number of zero modes \cite{tarnopolskyOriginMagicAngles2019a,beckerIntegrabilityChiralModel2022} that form a $C=1$ band, despite the lack of an external magnetic field. Indeed, the origin of P2 is not the index theorem \eqref{eq:index}, as $\dim \ker D = \dim \ker D^\dag$ for all $\alpha$. We conclude that there are examples of vortexable bands with no external magnetic field.

 The same relaxation mechanisms that led to suppressed $AA$ tunneling yields equal and opposite strain in both layers: $\boldsymbol{u}_+ = -\boldsymbol{u}_-$ \cite{koshinoEffectiveContinuumModel2020, carrMinimalModelLowenergy2019}. The resulting pseudomagnetic field $\mathcal{B}(\v{r}) = \nabla\times \mathcal{A}(\v{r})$ is moir\'e periodic, equal and opposite in the layers, and averages to zero over each moir\'e unit cell, but has peaks of ${\sim}80$\si{\tesla} \cite{koshinoEffectiveContinuumModel2020, carrMinimalModelLowenergy2019}. 
 
 It is interesting to examine the effect of this heterostrain on the vortexability of \eqref{eq:unstrained_chiral_limit}. We will see that while P2 is satisfied at a renormalized angle, P1 no longer holds. With strain effects taken into account, we obtain 
  \begin{equation}
      D_{\rm{TBG}}^{(\alpha_s)} = \begin{bmatrix}
      2 v\chi_+^\mu( -i\partial_\mu + \alpha_s \mathcal{A}_\mu) & w_1 U_1(\v{r})\\
      w_1 U_1(-\v{r}) & 2v \chi_-^\mu(-i \partial_\mu - \alpha_s \mathcal{A}_\mu)
      \end{bmatrix},
      \label{eq:strained_chiral_limit}
  \end{equation}
Property P2 remains satisfied. Indeed,
  Refs \cite{ShefferQueirozStern22,beckerSpectralCharacterizationMagic2021a,beckerFineStructureFlat2022} show the existence of a magic angle is robust to symmetry-preserving perturbations, which is indeed the case for realistic $\mathcal{A}(\v{r})$ \cite{koshinoEffectiveContinuumModel2020}. 
Furthermore, the position-dependent Fermi velocities, which lead to $\chi_\pm \neq \frac{1}{2} \begin{pmatrix} 1 & i \end{pmatrix}$, do not change the band energies. Indeed, we may work in the crystal frame of unperturbed atomic positions where $\chi^\mu \partial_\mu \to \ov{\partial}$. Thus, we still obtain an exactly flat band at a suitably renormalized $\theta(\alpha_s)$.
   We show in Fig. \ref{fig:internal_strain_vortexable_line} how the magic angle changes with $\alpha_s \mathcal{A}(\v{r})$, $0 \le \alpha_s \le 1$, giving rise to a ``magic line" of bands without any external magnetic field.

However, the different complex differential operators $\chi_\pm^\mu(\br) \partial_\mu$, associated with the opposite strains in both layers prevent us from defining a single vortex function $\sz$ that commutes with \eqref{eq:strained_chiral_limit} in the lab frame. Thus, property P1 is not satisfied. It is possible to define a generalized ``layer dependent vortex operator" $\tilde{\sz} = \diag(\sz_+, \sz_-)$ that satisfies $[D, \tilde{\sz}] = 0$ and $\tilde{\sz} \psi = \P \tilde{\sz} \psi$. But $\tilde{\sz}$ does not lead to a SRI-GS because on-site inter-layer repulsion is no longer screened: $\sz_+(\br) - \sz_-(\br) \neq 0$.
  
  We note that \eqref{eq:strained_chiral_limit} was recently studied in the final section of Ref. \cite{kangPseudomagneticFieldsParticlehole2022}, where a flat band also appeared. 
  In their chiral model, the authors of \cite{kangPseudomagneticFieldsParticlehole2022} work consistently to a lower order in gradients and hence do not include the position-dependent Fermi velocity, such that their chiral Hamiltonian may be interpreted as the crystal frame version of \eqref{eq:strained_chiral_limit}. With this approximation, \eqref{eq:strained_chiral_limit} is vortexable with $z=x+iy$ (and therefore hosts SRI-GS). 
  Note that previous non-chiral Hamiltonains in Refs. \cite{kangPseudomagneticFieldsParticlehole2022,VafekKangContinuumEffective2022} do include the position dependent Fermi velocity as well as other non-geometrical effects at the same order in gradients. Amongst the second order in gradient terms, we have kept only the geometric position-dependent Fermi velocity in order to study its effect on vortexability from a conceptual point of view.
  

\subsection*{E3: Degenerate Landau Levels in Chirally stacked Graphene (Multiple bands, total \texorpdfstring{$C>1$}{C>1}).}
For the next two examples we will focus on vortexable bands with higher Chern number. 
For simplicity, we will consider unstrained graphene and obtain standard vortex functions $z = x+iy$; strain may be included straightforwardly but here we want to emphasize the emergence of higher Chern numbers.

Recall that by ``band" we mean  ``low energy subspace with an extensive number of states"; vortexable bands can consist of multiple Bloch bands or Landau levels. In Bernal bilayer graphene it is well known that the zeroth and first Landau level are degenerate at zero energy \cite{novoselovUnconventionalQuantumHall2006,mccannElectronicPropertiesBilayer2013a}. This pattern extends to chirally-stacked graphene multilayers\cite{zhangSpontaneousQuantumHall2011,shiElectronicPhaseSeparation2020,geisenhofQuantumAnomalousHall2021} where Landau levels $0$ through $n-1$ are degenerate, where $n$ is the number of layers\cite{minChiralDecompositionElectronic2008}. To understand the vortexability of the \emph{combined} set of Landau levels, consider the Hamiltonian for AB stacked Bernal bilayer graphene in a constant magnetic field $B_0$:
 \begin{equation}
     H_{\text{Bernal-Bilayer}} = \begin{pmatrix} 0 & D^\dag_1 & 0 & 0 \\ D_1 & 0 & \gamma & 0 \\ 0 & 0 & 0 &  D^{\dag}_1 \\ 0 & 0 & D_1 & 0 \end{pmatrix}
     \label{eq:bernal_Bfield}.
 \end{equation}

In \eqref{eq:bernal_Bfield} the Hamiltonian acts on a wavefunction $\begin{pmatrix}\psi_{A1} & \psi_{B1} & \psi_{A2} & \psi_{B2} \end{pmatrix}^T$. The interlayer tunneling strength is $\gamma$, and 
\begin{equation}
  D_1 = 2v(-i\ov{\partial}_z - \ov{A}_z) = \sqrt{2} v\ell^{-1} \hat{a} 
\end{equation}
is the monolayer zero mode operator where $\ov{\partial}_z = \frac{1}{2}(\partial_x + i \partial_y)$, $\ov{A}_z = \frac{1}{2}(A_x + i A_y)$, $\hat{a}$ is the standard Landau level lowering operator, and $\ell = 1/\sqrt{B}$ is the magnetic length. We have neglected smaller interlayer hoppings \cite{jungAccurateTightbindingModels2014}.

Let us write \eqref{eq:bernal_Bfield} in the form \eqref{eq:chiral_ham_general}. We obtain
\begin{equation}
    D_{\text{Bernal-Bilayer}} = \begin{pmatrix} D_1 & \gamma \\ 0 & D_1 
    \end{pmatrix}.
    \label{eq:bernalzeromode}
\end{equation}
There are two types of zero modes of \eqref{eq:bernalzeromode} corresponding to a zeroth and first Landau level
\begin{equation}
   \psi_0 = \begin{pmatrix}
       \psi_{\text{LLL}} \\ 0 
   \end{pmatrix} 
   \qquad  \psi_1 = \begin{pmatrix}
       \hat{a}^\dag \psi_{\text{LLL}} \\ -\frac{\sqrt{2} v }{\ell \gamma} \psi_{\LLL}
   \end{pmatrix}.
   \label{eq:bernal_ZMs}
\end{equation}
Here $\psi_{\LLL}$ is a LLL state which satisfies $\hat{a} \psi_\LLL = \hat{a} z \psi_\LLL = 0$. The combined zero mode subspace \eqref{eq:bernal_ZMs}, consisting of the zeroth and first Landau level \emph{together}, is a vortexable band: the zero mode operator \eqref{eq:bernalzeromode} commutes with $z$. The space spanned by the zeroth Landau level $\psi_0$ is also vortexable on its own. However, the first Landau level on its own is not vortexable. Indeed, direct calculution shows that $z \psi_1$ has spectral weight in the zeroth Landau level, i.e. states of the form $\psi_0$ \eqref{eq:bernal_ZMs}, because $[a^\dag, z] \neq 0$. 

We may similarly write down the zero mode operator for $n$-layer chiral graphene which has the inductive form
\begin{equation}
    D_n = \begin{pmatrix}
        D_1 & \boldsymbol{\gamma}_{n-1} \\ 0 & D_{n-1}.
    \end{pmatrix}
    \label{eq:induction_chiral_graphene_Bfield}
\end{equation}
where $\boldsymbol{\gamma} = (\gamma, 0, \ldots, 0)$ is an $(n-1)$-dimensional hopping vector which we take to be nearest-neighbor for simplicity. The lowest $n$ Landau levels are therefore vortexable zero mode subspace of \eqref{eq:induction_chiral_graphene_Bfield}, with total Chern number $C = n$.

  \subsection*{E4: Chiral Graphene multilayers (\texorpdfstring{$C>1$}{C>1}, no magnetic field)}
  
  Our final example constructs vortexable bands with arbitrary Chern number $C>1$ without an external magnetic field. Our construction uses \textit{twisted} chiral graphene multilayers \cite{ledwithFamilyIdealChern2022,wangHierarchyIdealFlatbands2022a,liuGateTunableFractionalChern2021}:
  these are the structures above except the bottom layer is replaced by a chiral twisted bilayer graphene system. While this example is nominally that of a single physical system, we expect the general structure of this example to be a source of future zero-field higher Chern vortexable bands.

 Our zero mode operator is given by 
  \begin{equation}
      D_n = \begin{pmatrix} -2i v\ov{\partial}_z & \v{\gamma}_{n} \\ 0 & D_{n-1} \end{pmatrix}.
      \label{eq:inductive_ops}
  \end{equation}
  with $D_1 = D_{\textrm{TBG}}$ such that $D_n$ is associated with a chiral TBG system at its magic angle with $n-1$ monolayers stacked on top with zero twist angle such that each pair of monolayers is AB-stacked Bernal graphene. 

  The zero mode Bloch-periodic wavefunctions $\psi^{(n)}_{\bk,\ell}(\br)$ of $D_n$ on layer $1 \le \ell \le n$ may be obtained as
  \begin{equation}
      2iv\ov{\partial} \psi^{(n)}_{\bk,1}(\br) = \gamma \psi^{(n)}_{\bk, 2}(\br), \quad \psi^{(n)}_{\bk, \ell} = \lambda_\bk \psi^{(n-1)}_{\bk,\ell+1},
      \label{eq:inductive_soln}
  \end{equation}
  where $\lambda_\bk$ is $\bk$-dependent but otherwise constant. 

The first equation of \eqref{eq:inductive_soln} is generically solvable through Fourier series $\psi^{(n)}_{\bk, \ell < n}(\br) = e^{i \bk \cdot \br} \sum_\bG u^{(n)}_{\bk \ell<n}(\bG)  e^{i \bG \cdot \br}$ where $\bk$ is measured from the K point (for the $n$'th layer the periodicity is different due to the twist angle, though this will not play a role here \cite{ledwithFamilyIdealChern2022}). However, there is a single value $\bk = 0$ in the first Brillouin Zone for which there is an obstruction to inverting Eq. \eqref{eq:inductive_soln}. Indeed, we obtain
\begin{equation}
    -iv(k + G) u^{(n)}_{\bk,1}(\bG) = \gamma u^{(n)}_{\bk, 2}(\bG) = \gamma \lambda_\bk u^{(n-1)}_{\bk, 1}(\bG),
\end{equation}
where $k = k_x + i k_y$ and $G = G_x + i G_y$. Because $u^{(n-1)}_{\bk=-\bG, 1}(\bG) \neq 0$ generically, which we have verified numerically for this system, we must have $\lambda_{\bk = -\bG} = 0$ for all reciprocal lattice vectors $\bG$. 
  The additional phase winding associated with this zero leads to an increase in Chern number $C_n = C_{n-1} + 1$; see Ref. \cite{ledwithFamilyIdealChern2022} for details. Since the zero mode operator satisfies $[D_n, z] = 0$, we conclude that we have found a class of vortexable Bloch bands with arbitrary Chern number.

\section{Connection to momentum space band geometry}

We now specialize to the case of lattice translations, whereupon vortexability will become a computable \& quantifiable notion, and an organizing principle for many ideas in band geometry. ``Trace condition" bands are always \textit{special} vortexable bands, with a linear vortex function $\sz = x+iy$. We show the converse fails; the ``trace condition" is insufficient to identify \textit{general} vortexable bands where $\sz(\v{r})$ is nonlinear, such as E1 and E2 above. We then identify the point of failure --- the \textit{choice} $u_{\v{k}}(\v{r}) = e^{-i\v{k}\cdot\v{r}} \psi_{\v{k}}(\v{r})$ in Bloch's theorem. We end with a general formula, Eq. \eqref{eq:generalized_generalized_trace_condition}, that detects if a band is vortexable or nearly vortexable. Throughout the section we assume that the vortexable bands of interest satisfy the mild conditions in Sec. \ref{subsec:uniqueness}.

\subsection{Traditional Band Geometry and Vortexability with \texorpdfstring{$x+iy$}{x+iy}}

We first recall some standard definitions and results on band geometry. Consider a Chern $C$ band with an orthonormal basis $\ket{\psi_{\v{k}a}}$. Throughout this section we set $C, \Omega(\v{r}) \geq 0$ (the other case follows from complex conjugation). Momentum space band geometry characterizes the gauge-invariant part of the Bloch periodic wavefunctions $\ket{u_{\v{k}a}} := e^{-i\v{k}\cdot\v{r}} \ket{\psi_{\v{k}a}}$ through the Berry curvature $\mathcal{F}(\v{k})$ and Fubini-Study metric $g_{\mathrm{FS}}(\v{k})$ \cite{provost_riemannian_1980}. These are defined by (see e.g. \cite{parameswaranFractionalQuantumHall2013a,jacksonGeometricStabilityTopological2015,simonContrastingLatticeGeometry2020})
\begin{equation}
	g_{\mathrm{FS}}^{\mu \nu}(\bk) = \Re \tilde{\eta}^{\mu \nu}(\bk), \quad \F(\bk)  \varepsilon^{\mu \nu}= - \frac{1}{2}\Im \tilde{\eta}^{\mu \nu}(\bk),
	\label{eq:traditional_band_geometry}
\end{equation}
where 
\begin{equation}
    \tilde{\eta}^{\mu\nu}(\v{k}) = \sum_{a} \braket{\partial^{k_\nu} u_{\v{k}a} | Q_{\v{k}} | \partial^{k_\mu} u_{\v{k} a}}
    \label{eq:quantum_metric_main}
\end{equation}
 is the positive semidefinite quantum geometric tensor and $Q_{\v{k}} = I - \sum_{a} \ket{u_{\v{k}a}}\bra{u_{\v{k}a}}$. The lowest Landau level is entirely specified by its band geometry: any $C=1$ band where (I) the trace inequality
 \begin{equation}
\int  \tr g_{\mathrm{FS}}(\v{k})\;  d^2 \v{k} \ge 2\pi C
\label{eq:trace_condition}
\end{equation}
is saturated [the \textit{trace condition}] and (II) $\mathcal{F}(\v{k})$ is constant --- is equivalent to the LLL \cite{royBandGeometryFractional2014a}. The idea of ``Landau level mimicry" is that, so long as (I) and (II) are \textit{nearly} satisfied then, by adiabatic continuity, the band is likely to host a FCI phase. We reiterate that vortexable bands go beyond mimicry, hosting FQHE states in bands that do not resemble the LLL, with e.g. inhomogeneous Berry curvature or higher Chern number. Furthermore, in the SI (Sec. \ref{sec:band_geom_review}) we use our beyond-mimicry perspective to analyze other previously proposed band-geometric conditions such as the ``determinant condition", which we argue to be generically unrelated to the physics of the FQHE (Sec. \ref{subsec:detcond_folding}).

\begin{figure}
    \centering
    \includegraphics[width=\linewidth]{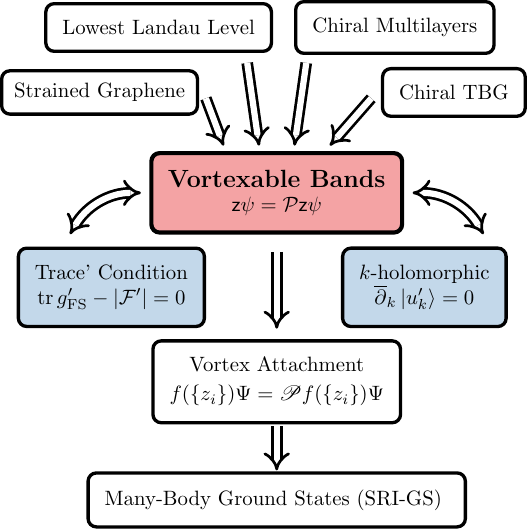}
    \caption{Flowchart describing examples, equivalent conditions, and consequences of vortexability. Several models, all describable using continuum chiral Hamiltonians \eqref{eq:chiral_ham_general}, lead to vortexable bands. Such bands satisfy the trace condition with a suitable choice of periodic wavefunction \eqref{eq:different_u} if the band is rotationally symmetric; if not a further linear transformation \eqref{eq:backtorotationallysymmetric} may be required. With this choice, and after a suitable non-unitary gauge transformation, the Bloch wavefunctions $u'$ can be chosen to be holomorphic in $k_x + i k_y$. Topologically nontrivial vortexable bands lead to \SRGS{}: FQHE ground states in the limit of short range interactions.}
    \label{fig:flowchart}
\end{figure}

Bands satisfy the trace condition if and only if they are vortexable with $\sz(\br) = x+iy$. 
Note that vortexibility involves multiplication of the $\psi$ wavefunctions by a position operator whereas band geometry consists of differentiating the $u$ wavefunctions with respect to $\bk$.
These settings are bridged by the product rule (as in Fourier transforms) through the definition $\ket{u_{\v{k}a}} := e^{-i\v{k}\cdot\v{r}} \ket{\psi_{\v{k}a}}$:
  \begin{equation}
    -i \bnabla_\bk \ket{\psi_\bk^a} = \hat{\br} \ket{\psi_\bk^a} +  e^{i \bk \br}  (-i\bnabla_\bk) \ket{u_\bk^a}.
      \label{eq:productrule_maintext}
  \end{equation}
  Acting with $\Q := 1-\P = 1-\sum_{\bk a} \ket{\psi_{\bk a}} \bra{\psi_{\bk a}}$
   annihilates the left hand side of Eq. \eqref{eq:productrule_maintext}, which can be seen by writing the $k$ derivative as a finite difference between wave-functions at two nearby $k$ points, and realizing that each of them belongs to the band and is hence annihilated by $\Q$. This gives:
  \begin{equation}
      \Q \br \ket{\psi_\bk^a} =   \Q  i e^{i \bk \cdot \br} \bnabla_\bk \ket{u_\bk^a}.
  \label{eq:projected_product_rule}
  \end{equation}
  For clarity, the projector $\Q$ acts on Bloch periodic wavefunctions $\ket{\psi_{\v{k}}}$ and contains a sum over $\bk$,  whereas $Q(\bk)$ acts on periodic wavefunctions, appears in the definition of quantum geometry \eqref{eq:quantum_metric_main}, and contains no sum over $\bk$. Since $\bra{\psi_{\bk'}} e^{i \bk \cdot \br} \nabla_\bk \ket{u_\bk}$ vanishes for $\bk \neq \bk'$ by translation symmetry, we have $\Q e^{i \bk \cdot \br} \bnabla_\bk \ket{u_{\bk}} = e^{i \bk \cdot \br} Q(\bk) \bnabla_\bk \ket{u_{\bk}}$, a rewriting of the right hand side of \eqref{eq:projected_product_rule}. Multiplication by $\v{\omega}^{(0)} =  \begin{pmatrix} 1 & i \end{pmatrix}^T$
  then yields
\begin{equation}
  \Q z \ket{\psi_{\bk a}} = 0 \iff Q(\bk) \ov{\partial}_k \ket{u_{\bk a}} = 0,
  \label{eq:trcond_vortexability_equivalence}
\end{equation}
where $z = x+iy$ and $2\ov{\partial}_k = \partial_{k_x} + i \partial_{k_y}$.  The right hand side of \eqref{eq:trcond_vortexability_equivalence} vanishes for all $\bk,a$ if and only if the trace condition is satisfied:
    \begin{align}
          \label{eq:trcond_holo}
      \int d^2 \bk (\tr g_\FS(\k) -  \F(\k)) & = \int d^2\bk\; \omega_\mu^{(0)} \tilde{\eta}^{\mu \nu}(\bk) \overline{\omega}_\mu^{(0)}
      \\
      \nonumber
      & = \sum_a \int d^2 \bk \; \norm{Q(\bk) \overline{\partial}_k u_\bk^a}^2.
      \end{align}
Thus a band is vortexable with $\sz = x+iy$ if and only if the trace condition is satisfied, and deviations can quantified using $\int \tr g_\FS d^2 \bk - 2\pi C \geq 0$.

A powerful alternative way to think about the trace condition is through $k$-space holomorphicity. The trace condition holds if, and only if, the Bloch wavefunctions $u_{\v{k}}$ can be written as holomorphic functions of $k = k_x + i k_y$ after a non-unitary gauge transformation $\tilde{u}_{\sk} = S_{\v{k}} u_{\v{k}}$ \cite{meraKahlerGeometryChern2021}. Here $S_{\v{k}}$ is an invertible matrix at each $\v{k}$, and acts by matrix multiplication in band indices. This claim is proven in the SI (see Prop \ref{prop:holo_gauge_choice} and Sec. \ref{subsec:holo_Bloch}). The key point is that the right-hand side of Eq. \eqref{eq:trcond_vortexability_equivalence} is a gauge-invariant version of the Cauchy-Riemann equations. If $Q(\bk) \ov{\partial}_k \ket{u_{\bk a}} = 0$, one may always solve $P(\bk) \ov{\partial}_k (S_\bk \ket{u_\bk}) = 0$ to fix a $k$-holomorphic gauge. Brillouin zone holomorphicity provides exceptionally strong constraints on the single-particle wavefunctions of vortexable bands, and is a key tool in our uniqueness result Prop. \ref{prop:uniqueness}.

The structures we have described in this subsection is a special case, corresponding to $\sz=x+iy$, of the interrelations depicted in Fig. \ref{fig:flowchart}. We will now cover the general case, with arbitrary $\sz$ and a suitably modified trace condition.

\subsection{Bloch's Theorem and General Vortexability}

\textit{General vortexable bands} are vortexable bands where $\sz(\v{r})$ is a nonlinear function of $\v{r}$. In light of examples E1 and E2 this class is physically relevant, leading to a natural question: given a band, how can one compute if it is a generally vortexable band? We must go beyond the trace condition Eq. \eqref{eq:trace_condition} which is equivalent to vortexability with $z = x+iy$.  Traditional band geometry \eqref{eq:traditional_band_geometry} can only detect linear vortex functions (special vortexability). The limitation is the single $\bk$-derivative acting on $\ket{u_\bk}$ in \eqref{eq:traditional_band_geometry}, which translates to a single power of the position operator $\br$ through the differentiation of $u_\bk(\br) = e^{-i\bk\cdot \br} \psi_\bk(\br)$ [see e.g. \eqref{eq:productrule_maintext}]. Thus we can only have an equivalence \eqref{eq:trcond_vortexability_equivalence} for linear vortex functions for this definition of $\ket{u_{\bk a}}$. 

However, we can instead define 
\begin{equation}
    \ket{u'_{\bk a}} = e^{-i \bk \cdot \br'} \ket{\psi_{\bk a}}
    \label{eq:different_u}
\end{equation}
where $\br'(\br+\ba) = \br'(\br) + \ba$. This, in turn, defines $\mathcal{F}'$ and $g_{FS}'$. The reasoning above \eqref{eq:trcond_vortexability_equivalence} then yields
\begin{equation}
  \Q z' \ket{\psi_{\bk a}} = 0 \iff Q(\bk) \ov{\partial}_k \ket{u'_{\bk a}} = 0,
  \label{eq:trace_vortex_equiv_afterdiffeo}
\end{equation}
such that the trace condition \eqref{eq:trace_condition}, defined using the wavefunctions $\ket{u'_{\bk a}}$ is equivalent to vortexability with $\sz = z' = x' + i y'$.
We underscore that \eqref{eq:different_u} is \textit{not} related to the usual definition by a gauge transformation. Indeed, such a transformation is akin to changing unit cell embedding in tight-binding models and yields different values for gauge-invariant quantities such as the Berry curvature: $\mathcal{F}_{\v{k}}' \neq \mathcal{F}_{\v{k}}$ \cite{simonContrastingLatticeGeometry2020}. We note that the traditional choice $\ket{u_{\bk a}} = e^{-i \bk \cdot \br} \ket{\psi_{\bk a}}$ is often convenient; for example electric fields couple via $\v{E} \cdot\v{r}$ such that the traditional $\ket{u_{\v{k}}}$ appears in semiclassical transport \cite{simonContrastingLatticeGeometry2020}. For our purposes, however, widening our scope to the modification \eqref{eq:different_u} is essential.

We now show how \eqref{eq:different_u} and \eqref{eq:trace_vortex_equiv_afterdiffeo} characterize general vortexable bands, starting with the case of rotational symmetry for simplicity. Any vortex function may be written as $\sz(\br) = x^{(\sz)} + i y^{(\sz)}$ where $\br^\sz(\br): \mathbb{R}^2 \to \mathbb{R}^2$ is a diffeomorphism [see below Eq. \eqref{eq:non-degeneracy_condition}]. With $n>2$-fold rotation symmetry, vortex functions obey $\sz(\br + \ba) = \sz(\br) + a_x + i a_y$ under translations (see SI Prop. \ref{prop:rotation_symmetry_constraints}). 
Then $\br^{(\sz)}(\br + \ba) = \br^{(\sz)}(\br) + \ba$ such that the wavefunctions $\ket{u^{(\sz)}_{\bk a}} = e^{-i \bk \cdot \br^{(\sz)}} \ket{\psi_{\bk a}}$ are periodic and \eqref{eq:trace_vortex_equiv_afterdiffeo} is satisfied for $\sz = x^{(\sz)} + i y^{(\sz)}$. We define

\begin{equation}
      T = \left [ \min_{\br'(\br)} \int 
 \tr g'_{\FS} \; d^2 \bk \right ] - 2\pi C \ge 0
        \label{eq:generalized_trace_condition}
\end{equation}
where $\v{r}'$ runs over lattice-periodic, orientation-preserving diffeomorphisms. A ($C_{n\ge 2}$ symmetric) band is vortexable (with $C, \Omega(\v{r}) \ge 0$) if and only if $T=0$. The minimizing diffeomorphism $\v{r}' = (x',y')$ gives the vortex function $\sz = x' + i y'$.

The modification \eqref{eq:generalized_trace_condition} of band geometry was previously proposed in Ref. \cite{simonContrastingLatticeGeometry2020} in the tight-binding setting (see also \cite{huhtinen2022revisiting} for $C=0$). There it was interpreted as a choice of unit cell embedding in tight binding models. Here it emerges naturally for continuum vortexable bands in terms of the choice of periodic wavefunction \eqref{eq:different_u}: this is precisely where the choice of coordinate frame or unit cell embedding enters band geometry.

	For bands without rotational symmetry, we must modify Eq. \eqref{eq:generalized_trace_condition} slightly. In general (see SI Props \ref{prop:general_translation} \& \ref{prop:holo_gauge_choice}),
\begin{equation}
  \sz(\v{r}+\v{a}) = \sz(\v{r}) + \omega_\mu a^\mu,  
  \label{eq:generic_translation_maintext}
\end{equation} 
 for some complex vector $\v{\omega}  =  \v{\omega}^{(0)} J$ where $\v{\omega}^{(0)} = \begin{pmatrix} 1 & i \end{pmatrix}^T$ and $J$ is a $2 \times 2$ invertible matrix. 
 Note that the transformation
 \begin{equation}
   \br \to J^{-1} \br, \quad \ba \to J^{-1} \ba, \quad  \bk \to \bk J,
   \label{eq:backtorotationallysymmetric}
\end{equation}
reduces us to the rotationally-symmetric case. Therefore we require a form of the trace condition that is invariant under linear transformations.

The reformulation of the trace condition we need is the zero mode condition $\omega_\mu \tilde{\eta}^{\mu \nu}(\bk) = 0$ for $\v{k}$-independent  $\omega_\mu$ \cite{wangChiralApproximationTwisted2021a}.
 It is invariant under linear transformations induced by \eqref{eq:backtorotationallysymmetric}, $\v \omega \to \v \omega J^{-1}$ and $\tilde{\eta} \to J \eta J^T$, and it reduces to the the usual trace condition 
 $\omega^{(0)}_\mu \tilde{\eta}^{\mu \nu} = 0$ [see \eqref{eq:trcond_holo}]
 under a suitable choice of $J$. 
Such a zero mode $\omega_\mu$ exists if and only if $\det \tilde{\eta}_I = 0$ with 
\begin{equation}
    \tilde{\eta}_I = \int_{\text{BZ}} \tilde{\eta}(\v{k}) \, d^2\v{k},
\end{equation}
 because $\tilde{\eta}(\bk)$ is positive definite for each $\bk$. Let us then define
\begin{equation}
      T = \min_{\br'(\br)} 2\sqrt{\det \tilde{\eta}_I} \ge 0,
        \label{eq:generalized_generalized_trace_condition}
\end{equation}
such that a band with discrete translation symmetry is vortexable with a translationally symmetric vortex function $\partial_\mu \sz(\br) = \partial_\mu \sz(\br + \ba)$ if and only if $T = 0$ in \eqref{eq:generalized_generalized_trace_condition}. The vortex function is given as $\sz = \omega_\mu r'^\mu$ where $\br'(\br)$ is the minimizing diffeomorphism of \eqref{eq:generalized_generalized_trace_condition} and $\omega_\mu$ is the zero mode that leads to the vanishing of $\det \tilde{\eta}_I$. We have included a square root and factor of $2$ such that \eqref{eq:generalized_generalized_trace_condition} reduces to \eqref{eq:generalized_trace_condition} in the rotationally symmetric case.
It would be interesting to study the regime of ``nearly" vortexable bands where $T$ is small but non-zero.

\section{Conclusions}

In this work we have introduced vortexable bands: bands which admit vortex attachment through a complex valued function $\sz(\br)$. These bands allow construction of \SRGS --- exact many-body FQHE ground states in the limit of short-range interactions. Vortexable bands emerge naturally in graphene-based systems with or without magnetic field, with arbitrary Chern number, and with arbitrary vortex functions $\sz(\br)$. A subset of ``special" vortexable bands, with $\sz = x+iy$ and discrete translation symmetry, may be characterized using traditional approaches to momentum space band geometry, in particular the trace condition. By moving beyond the conventional choice of periodic wavefunctions $u_\bk(\br) = e^{-i \bk \cdot \br} \psi_\bk(\br)$ we arrived at a formula that can check if any band is vortexable.

Our treatment of vortexable bands 
opens up a number of new directions for future investigation. What types of vortexable bands are possible? Can we characterize the single particle wavefunctions? Can we define the real-space geometry of general bands, such that the Kahler geometry \eqref{eq:vortex_geometry} emerges when the band is vortexable? How can we understand the interplay between interaction range and Berry curvature inhomogeneity?
What types of FCIs do they host, especially with higher Chern number? How do composite Fermi liquids or non-Abelian states behave in generic vortexable bands? What happens to excitations like Skyrmions or magneto-roton modes in vortexable bands that are dissimilar to the LLL? Chern bands with small dispersion and small $T$ [Eq. \eqref{eq:generalized_trace_condition}] should host FCIs --- can one derive a quantitative relation to the many-body gap? Are vortexable bands of equal and opposite Chern numbers ideal for other phases such as superconductors? We believe that future work will focus on the above questions and more. 

More broadly, vortexable bands combine the analytic tractability of the LLL with much of the diversity of topological bands in condensed matter systems (in particular, magnetic translation symmetry and flat Berry curvature is not required). We therefore argue that vortexability will provide a central conceptual role in the rich and emerging field of interacting topological bands.

\section{Acknowledgements}

We would like to thank Ruihua Fan, Ilya Esterlis, Rahul Sahay for thorough comments and suggestions that greatly improved the manuscript. We would also like to thank Eslam Khalaf, Junkai Dong, Jong Yeon Lee, Bertrand Halperin, Saranesh Prembabu, Gregory Parker, and Jie Wang for helpful discussions and collaborations on related topics. 

A.V. was supported by a
Simons Investigator award by the Simons Collaboration on Ultra-Quantum Matter, which is a grant from
the Simons Foundation (651440, AV) and by NSF-DMR 2220703. P.J.L. was supported by the Department of Defense (DoD) through the
National Defense Science and Engineering Graduate Fellowship (NDSEG) Program.
This research is funded in part by the Gordon and Betty
Moore Foundation’s EPiQS Initiative, Grant GBMF8683
to D.E.P.

\appendix

\section{Proof of $\Omega C \geq 0$ and comments on $C=0$ vortexability
    } \label{sec:chirality_chern}
    In this section we prove that the chirality $\Omega$ of vortex geometry is tied to the Chern number of the band and then use the result to understand $C=0$ vortexable bands. There are several ways to show this result, including momentum space band geometry for translationally symmetric systems --- which is done in the final section of the main text --- as well as the closely related twist angle torus geometry \cite{niuQuantizedHallConductance1985,watanabeInsensitivityBulkProperties2018} for systems that are placed on a twist-angle torus with arbitrary twisted boundary conditions. Here, however, we will take a bulk position operator point of view\footnote{We would like to thank Ruihua Fan for patient help in understanding this perspective, both conceptually and computationally.} and use the formula \cite{bellissardNonCommutativeGeometryQuantum1994,prodanDisorderedTopologicalInsulators2011,kitaevAnyonsExactlySolved2006}
  \begin{equation}
    C = 2\pi i \Tr' [\P x' \P, \P y' \P].
    \label{eq:position_operator_chern}
  \end{equation}
  where $\Tr'$ is the trace per unit volume
  \begin{equation}
      \Tr' A = \lim_{\abs{\Lambda} \to \infty} \frac{1}{\abs{\Lambda}} \tr P_\Lambda A P_\Lambda
      \label{eq:trace_per_unit_volume}
  \end{equation}
  where $\Lambda$ labels successively larger regions of real space, $P_\Lambda$ projects onto states that vanish outside $\Lambda$, and we use the volume form $dx' \wedge dy'$ associated with the coordinates $(x',y')$ to measure the volume $\abs{\Lambda} = \int_{\Lambda} dx' \wedge dy'$. The trace $\tr$ is the usual dimensionless Hilbert-space trace.
  Note that some caution is required in using \eqref{eq:trace_per_unit_volume} since the trace is over an infinite dimensional space. In particular, the cyclic property only holds when the trace exists and cannot be used on each term of the commutator \eqref{eq:position_operator_chern} \emph{individually}.
  For systems with lattice translation symmetry  we recover the usual formula from \eqref{eq:position_operator_chern}. In particular we have
  \begin{equation}
      [\P x' \P, \P y' \P] = -i \sum_{\bk} \F'(\bk) \ket{\psi_{\bk}} \bra{\psi_{\bk}}, 
  \end{equation}
  such that
\begin{equation}
      \Tr' [\P x' \P, \P y' \P] = -i\frac{1}{V} \sum_\bk \F'(\bk) = -i\int \frac{d^2 \bk}{(2\pi)^2} \F'(\bk),
\end{equation}
  where the Berry curvature $\F'$ is defined using the $(x',y')$ coordinate system and $V = \int dx' \wedge dy'$. For a system on a twist-angle torus we may also use a similar calculation to show that \eqref{eq:position_operator_chern} reproduces the expected result\footnote{Twist angle boundary conditions are formally equivalent to Bloch boundary conditions if one chooses the entire system to be a single unit cell}.
   
   The virtue of \eqref{eq:position_operator_chern} is that it makes sense on the infinite plane and is stated in terms of the position operator.
  We use the coordinates $(x', y')$ because we will soon plug in the vortex function $\sz = x' \pm i y'$. 
  As a topological quantity, the Chern number is invariant under of choice of coordinates --- though this is not always true for the Hall conductance in states without a charge gap \cite{simonContrastingLatticeGeometry2020}.

  See Refs. \cite{bellissardNonCommutativeGeometryQuantum1994,prodanDisorderedTopologicalInsulators2011,kitaevAnyonsExactlySolved2006} for a careful discussion of Eq. \ref{eq:position_operator_chern}. Note that in these references the trace is often written in the equivalent form $\Tr' \P[[x,\P][y,\P]] = \Tr' \P\left[ [x,\P],[y,\P] \right] \P$ to emphasize the role of the position operator as a generalization of a derivative.

  We now take $\Omega > 0$ such that $\sz = x'+iy'$ and show that $C \geq 0$. The analogous case of $\Omega < 0$ follows straightforwardly. We now compute
  \begin{equation}
  \begin{aligned}
    C & = \pi \Tr' [\P \ov{\sz} \P, \P \sz \P] = \pi \Tr'( \P \ov{\sz}\sz \P - \P \sz \P \ov{\sz}\P) \\
    & = \pi \Tr' \P \sz (1-\P) \ov{\sz} \P = \pi \Tr' \P \sz \Q \ov{\sz} \P \geq 0. 
    \end{aligned}
\label{eq:expand_commutator}
  \end{equation}
  where we expanded the commutator, used the vortexability condition $\P \sz \P =  \sz \P$ in the first term, and used the commutativity $[\ov{\sz}, \sz]=0$ of the unprojected vortex functions. The non-negativity comes from the writing $\P \ov{\sz} \Q \sz \P = A^\dag A$ as a positive definite operator where $A = \Q \sz \P$. 

  \subsection{Vortexable bands with $C=0$}\label{subsec:chern_zero}
  Using the above results, we now shows that $C=0$ vortexable bands are ``trivial" in the sense that they do not support FQHE ground states.
  
   If $C=0$, then by \eqref{eq:expand_commutator}, $C = \pi \Tr A^\dag A = \pi \norm{A}^2 = 0$, so we must have $A = \Q \sz \P = 0$. Thus, if $C=0$ then the band is \emph{also} vortexable with vortex operator $\ov{\sz}$. This makes some sense, as $C=0$ bands intuitively don't have have a bias towards either chirality. However, this dramatically restricts the wavefunctions of the $C=0$ band. For bands with translation symmetry, for example, it is possible to show that the wavefunctions $\ket{u_\bk}$ may be chosen constant in $\bk$ such that $\F = g_\FS = 0$. However, we will instead proceed by characterizing the wavefunctions without reference to the translation symmetric case.
   
   Indeed, since the vortexability condition is linear we have $(\sz \pm \ov{\sz}) \psi = \P (\sz \pm \ov{\sz}) \psi$ such that 
  \begin{equation}
    \br' \psi = \P \br' \psi.
    \label{eq:pos_operator_attach}
  \end{equation}
 
  We may extend this to any analytic function of $\br'$ by expanding in a Taylor series. This seems to enable us to construct an infinite number of states in any finite volume. Indeed, consider the attachment
  \begin{equation}
    f_{\br'_0}(\br') \ket{\psi} = \P f_{\br'_0}(\br') \ket{\psi}, \,\,\, f_{\br'_0}(\br) = \frac{1}{2\pi a^2}e^{-\frac{\norm{\br' - \br'_0}^2}{2a^2}},
    \label{eq:localize_attach}
  \end{equation}
  Which localizes the wavefunction $\ket{\psi}$ around some position $\br_0$ where $\ket{\psi}$ has nonzero density with localization length $a$. Performing \eqref{eq:localize_attach} for arbitrarily small $a$ and arbitrarily many positions $\br'_0$ seems to enable the construction of an arbitrarily large number of states in a finite volume. The one loophole is that $\ket{\psi}$, and all other states in the band of interest, may have only finitely many points $\br'_0$ with nonzero density, such that the vast majority of attachments \eqref{eq:localize_attach} simply give zero. This is the case in the tight-binding limit for example, but we find it remarkable that all $C=0$ vortexable bands with finite fully-filled density are \emph{exactly} in this limit of completely localized states. 

  We now show that we can describe such bands using a basis of position eigenstates; i.e. the band admits a basis of \emph{delta-function localized} ``Wannier orbitals". Applying \eqref{eq:pos_operator_attach} to a basis of states in the band we obtain
  \begin{equation}
      \br' \P = \P \br' \P
  \end{equation}
  and the Hermitian conjugate $\P\br' = \P \br'\P$. We therefore conclude
  \begin{equation}
      [\br',\P] = 0
  \end{equation}
  such that the position operator and the band projector are simultaneously diagonalizable. The band therefore admits a basis in terms of the eigenstates of the position operator as claimed: $\hat{\br}' \ket{\psi_\alpha} = \br'_\alpha \ket{\psi_\alpha}$.
  
  To minimize repulsive interaction energy at fractional filling, we may simply occupy some subset of the orbitals $\ket{\psi_\alpha}$ and maximize the minimum distance between occupied orbitals. For instance, one can use a charge-density wave pattern. If, for a given filling factor, the minimum distance between occupied orbitals is $d$, then the state will be a zero energy ground state of any potential with $V(\norm{\br}) = 0$ for $\norm{\br} > d$. This is a much larger class of interactions than the pseudopotential Hamiltonians, which are all ultra-local and only make sense in continuum models. 

  It is also worth noting that if we start with the fully filled $C=0$ Slater determinant state, then the vortex attachment \eqref{eq:generalized_Laughlin} simply multiplies the wavefunction by a constant. Indeed, the fully filled Slater determinant may be written as the antisymmetric tensor product of position eigenstates such that the vortex operators $\sz_i$ get replaced by their eigenvalue.

  We therefore conclude that vortex attachment in $C=0$ vortexable bands is ineffective in producing fractional quantum Hall states; charge density waves are much more effective at minimizing the interaction energy due to the perfect localization of the single particle states.

    \section{Pseudopotential expansions}\label{sec:pseudo}

In this section we argue that the Laughlin-like state
    \begin{equation}
      \ket{\Psi^{(2s)}} = \prod_{i<j} (\sz_i - \sz_j)^{2s} \ket{\Psi}
    \label{eq:generalized_Laughlin_supp}
    \end{equation}
    is the ground state for $C \neq 0$ continuum vortexable bands in the limit of short-range interaction potentials. That is, it is an ``SRI-GS".  We follow the argument in Ref. \cite{trugmanExactResultsFractional1985a}, which used a real-space version of the Haldane pseudopotentials \cite{haldaneFractionalQuantizationHall1983a} to argue that the Laughlin state does not crucially depend on angular momentum or rotation symmetry. This argument has recently been applied to chiral twisted bilayer graphene as well \cite{ledwithFractionalChernInsulator2020a,ledwithStrongCouplingTheory2021,wangExactLandauLevel2021a}.

  We consider a purely-interacting Hamiltonian projected to the band of interest
    \begin{equation}
      H = \Pmany \sum_{i<j} V\left(\br_i, \br_j \right) \Pmany
      \label{Ham}
    \end{equation}
    where $\Pmany = \otimes_i \mathcal{P}_i$ is the many-body projector to the band of interest.  We take $V(\br,\br') = V(|\br-\br'|)  > 0$ to be a circularly and translationally symmetric repulsive interaction potential --- both for simplicity and because we have a screened-Coulomb-type interaction in mind. 
    The essential piece of the argument below will be power-counting in the short range limit, which may therefore be generalized to any short range repulsive interaction.

 We begin by expanding the interaction potential into ``pseudopotentials", written in a real space basis. To understand this expansion, we note that the potential $V(\br) = V(r)$ will always be integrated against some matrix element or probability-density of states in the band (for the energy, this will be ultimately be the pair distribution function). Call this function $\Phi(\br)$. The function $\Phi(\br)$ generically varies on the scale of the effective magnetic length $\ell$ where $2 \pi \ell^{-2}$ is the change in electron density upon fully filling the band. An important exception to this is the tight-binding limit where electrons have a finite average density but are taken to be infinitely localized - this discussion will therefore apply most saliently to continuum models. Note that the tight binding limit is itself an approximation, however, and real electron atomic orbitals tend to have appreciable spread: of the same order as the inter-site distance. We will now proceed assuming $\Phi$ varies on the scale of $\ell$.
  \begin{equation}
      \int d^2 \,\br V(\br) \Phi(\br).
  \end{equation}
  
While $\Phi$ is usually not circularly symmetric \textit{a priori}, we may assume it to be so because it is integrated against $V(\br) = V(r)$. 
  Motivated by Haldane pseudopotentials for short-range interactions, we expand near $\br = 0$ using the Taylor-like expansion for circularly symmetric analytic functions
  \begin{equation}
      \Phi(\br) = \sum_{n=0}^\infty c_n^{-1}(\nabla^{2n} \Phi)(\br = 0) r^{2n},
  \end{equation}
  where $c_n = (\nabla^{2n} r^{2n})|_{r=0} = 4^n(n!)^2$. We have
  \begin{equation}
  \begin{aligned}
      & \int d^2 \br\, V(\br) \Phi(\br) \\
      & = \sum_{n=0}^\infty c_n^{-1} ((\ell\nabla)^{2n} \Phi)(0) \int d^2 \br \left(\frac{r}{a_M}\right)^{2n}V(\br),
      \end{aligned}
  \end{equation}
  where we have inserted factors of $\ell$ to effectively non-dimensionalize $r$ because $\Phi$ varies on the scale of $\ell$. 

  The above result may be interpreted as coming from a ``pseudopotential" expansion
  \begin{equation}
  \begin{aligned}
      V(\br) & = \sum_{n=0}^{\infty} v_n (\ell \nabla)^{2n} \delta(\br) \\
      v_n & = \frac{1}{c_n}\int d^2 \br \left(\frac{r}{\ell}\right)^{2n} V(\br)  \sim \left(\frac{d}{\ell}\right)^{2n} v_0,
      \end{aligned}
\label{eq:pseudo_expand_long}
  \end{equation}
  through the use of integration by parts. Here, $d$ is the range of the interaction. If $d \ll \ell$ the coefficients $v_n$ rapidly decrease with $n$, though they may decrease rapidly regardless due to the extremely rapid growth of $c_n = 4^n (n!)^2$.

  \subsection{Ground state in the pseudopotential limit}

  We now show that the Laughlin-like state $\ket{\Psi^{(2s)}}$, defined in \eqref{eq:generalized_Laughlin_supp} through attaching $2s$ vortices, is a zero mode under the first $2s-1$ terms of \eqref{eq:pseudo_expand_long} such that the ground state energy scales to zero no-slower-than $d^{4s}$ as $d \to 0$. 
  \footnote{In the ordinary spin-polarized FQHE an additional vortex from Fermi statistics makes the scaling $d^{4s+2}$ instead, for example. This extra enhancement is not always present. For example it is absent in the spin-singlet $(2s+1,2s+1,2s)$ Halperin states which also arise from vortex attachment.}

  The energy expectation value is
  \begin{equation}
  \begin{aligned}
    E & = \sum_{i<j} \int  V(\br_i - \br_j) d^2 \br_i  d^2 \br_j  \\
    & \qquad \times \int \norm{\Psi^{(2s)}(\br_1, \ldots, \br_N)}^2 \prod_{k \neq i,j} d^2 \br_k  \\
      & = N \int d^2 \br d^2 \br_0 V(\br - \br_0) G_{\br_0}(\br - \br_0),
    \end{aligned}
    \label{eq:energy_expectation_generic}
  \end{equation}
  where
  \begin{equation}
  \begin{aligned}
    & G^{(2s)}_{\br_0}(\br_-) \\
    & = (N-1) \int \prod_{k>2} d^2 \br_k \norm{\Psi^{(2s)}(\br_0, \br_0 + \br_-, \br_3,  \ldots, \br_N)}^2 
    \end{aligned}
    \label{eq:pair_dist}
  \end{equation}
  is the pair distribution function as a function of the relative coordinate $\br_- = \br - \br_0$. Above we used that $\norm{\Psi_{2s}(\{\v{r}_i\})}^2$ is a symmetric under permuting the positions of the particles. The vortex attachment \eqref{eq:generalized_Laughlin} implies that 
  \begin{equation}
    G^{(2s)}_{\br_0}(\br_-) = (g^v_{\mu \nu}(\br_0) r_-^\mu r_-^\nu)^{4s}\lambda_{\br_0}(\br_-) + \ldots
    \label{eq:pair_dist_behavior}
  \end{equation}
  in terms of the vortex metric $g(\v{r}_0)$ defined in the main text. Here the dotted terms vanish with even higher powers of $r_-^\mu$ and $\lambda_{\br_0}(\br_-)$ is some function that we will not be concerned with here, though we ask that it is finite (i.e. not infinite, it is ok if it vanishes) as we would expect in a continuum model with a smooth $G_{\br_0}(\br_-)$.

 Using the psuedopotential expansion \eqref{eq:pseudo_expand_long},  the energy is then given by
  \begin{equation}
  \begin{aligned}
    E = \sum_{n = 0}^\infty N \int d^2 \br & d^2 \br_0  v_n  (\ell \nabla)^{2n}\delta(\br_-) \\
    & \times\left( (g_{\mu \nu}(\br_0) r_-^\mu r_-^\nu)^{4s}\lambda(\br_0) + \ldots \right).
    \label{eq:finalE}
    \end{aligned}
  \end{equation}
  We then move all derivatives off the delta function through integration by parts, and evaluate the rest of the integrand at $\br_- = 0$ in accordance with the delta function. We see that for $n < 2s$ there are more powers of $r_-^\mu$ than derivatives such that the integrand must vanish at $\br_- = 0$. Thus, the energy only receives contributions from $n \geq 2s$ such that the ground state energy is expected to scale as $d^{4s}$ for small $d \ll \ell$. 
 We expect that other competing states will not be zero modes under all terms with $n<2s$, such that the state \eqref{eq:generalized_Laughlin_supp} emerges as the ground state as $d \to 0$.

\section{Strain, Graphene, and Twisted Bilayer Graphene}
\label{app:strain_graphene_TBG}

This Appendix gives details of strain in graphene and TBG that are used in examples in the main text.

\begin{figure*}
    \centering
    \includegraphics[width=\textwidth]{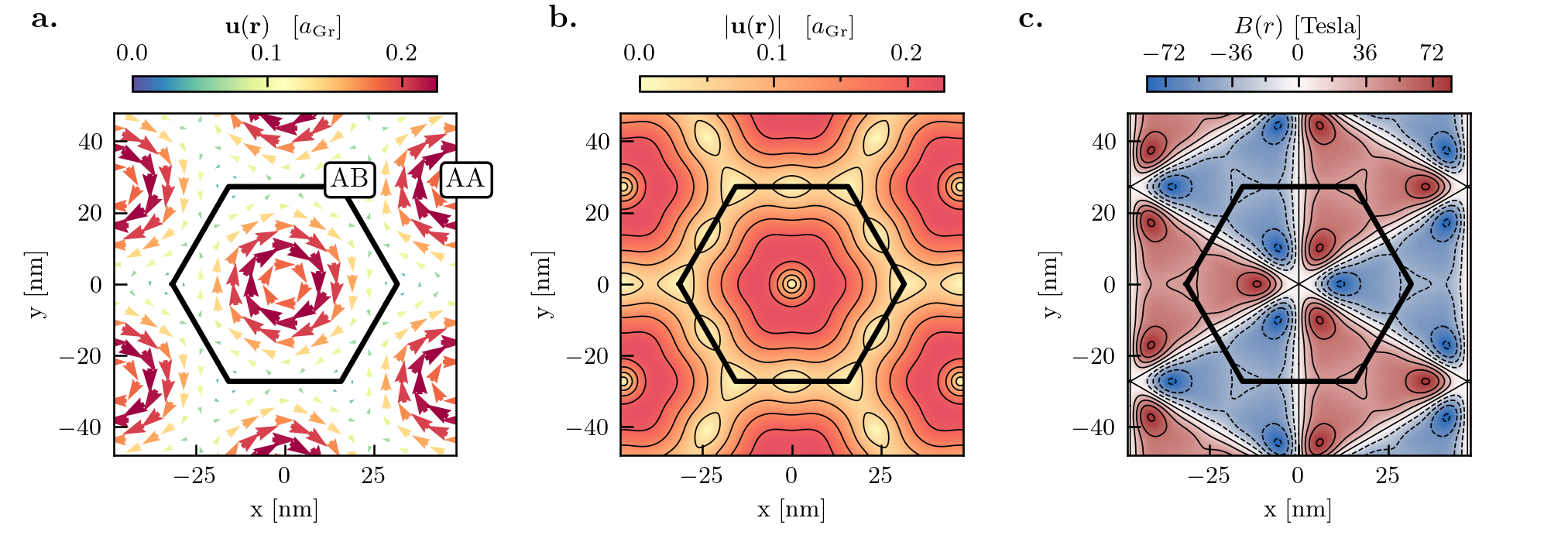}
    \caption{Internal strain of TBG. a. Displacement field $\v{u}(\v{r})$ in units of the graphene scale $a_{\mathrm{Gr}}$. The solid hexagon is a real-space unit cell. b. Magnitude of the displacement field $\n{\v{u}(\v{r})}$. c. Psuedomagnetic field in real space. All data is for $\theta=1.05^\circ$ and uses \textit{ab initio} strain parameters from \cite{koshinoEffectiveContinuumModel2020}.}
    \label{fig:strained_TBG}
\end{figure*}

\subsection{Graphene}

To set conventions, let $\v{R}_{1,2} = (1/2,\mp\sqrt{3}/2) a$ be the lattice of graphene, where $a$ is the graphene lattice constant. Consider graphene where atoms are displaced from their equilibrium positions by $\v{u}(\v{r})$. One can think of this as embedding a 2d crystal of graphene --- with regularly spaced atoms --- into 3d space in the laboratory. These two frames are called the \textit{crystal} and \textit{laboratory} frames respectively, and have coordinates $\v{r}'$ and $\v{r}$ \cite{de2012space,de2013gauge,manesGeneralizedEffectiveHamiltonian2013}. The latter must be used when coupling to all external probes. For instance, electron fields couple as $\v{E}\cdot \v{r}$. However, the lattice is only evenly spaced in the crystal frame, making it useful for Fourier transforms. The metric of the crystal frame is trivial, but the embedding makes the metric non-trivial in the laboratory frame:
\begin{equation}
	g_{\mu\nu}(\v{r}) = \delta_{\mu\nu} + 2 u_{\mu\nu}(\v{r}) = e_\mu^a(\v{r}) e_\nu^b(\v{r}); 
\end{equation}
where $ e_{a \mu}(\v{r}) = \delta_{a \mu} + \delta_a^\nu u_{\mu\nu}$ are tetrads giving an local orthonormal basis \cite{de2012space} and $\quad u_{\mu\nu}(\v{r}) = \frac{1}{2}[ \partial_\mu u_\nu + \partial_\nu u_\mu]$ is the strain tensor in terms of the atomic displacements $u_\mu(\br)$.

At first order in a derivative expansion, the low energy continuum model of strained graphene is:
\begin{equation}
\begin{aligned}
	\hat{h}_{\text{strained graphene}} & =  v_F \int d^2 \v{r} \hat{\psi}^\dagger e_a^\mu \sigma^a (-i \partial_\mu - \mathcal{A}_\mu(\v{r})) \hat{\psi}(\v{r}), \\
  \mathcal{A} & = \frac{\beta}{2a} (u_{xx}-  u_{yy}, -2u_{xy}),
	\label{eq:strained_graphene_appendix}
 \end{aligned}
\end{equation}
where the strain gauge field $\v{\mathcal{A}}$ is determined by the elastic coefficient $\beta = \partial \ln t(d)/\partial \ln d\big|_{d=a}$ in terms of the hopping matrix element $t$ at separation $d$, and $v_F$ is the Fermi velocity \cite{neto2009electronic}. Remarkably, the Hamiltonian takes the form \eqref{eq:strained_graphene_appendix} in both frames at this order in derivatives, though it differs at second order in derivatives, where a spin connection enters \cite{de2013gauge}.

\subsection{Strained Twisted Bilayer Graphene}
At twist angle $\theta$, the unstrained moir\'e reciprocal lattice is $\v{g}_{1,2} = (2\pi/\sqrt{3}, \pm 2\pi) L_M^{-1}$ where $L_{M} = [2 \sin \frac{\theta}{2}]^{-1} a$ in terms of the graphene unit cell. A standard model for the low energy Hamiltonian (in the $K$-valley) is \cite{ledwithStrongCouplingTheory2021}:
\begin{equation}
\begin{aligned}
	\hat{H}_{\mathrm{TBG}}(\v{r}) & = \begin{pmatrix} 
		-i v_F \v{\sigma} \cdot \v{\nabla} & T(\v{r})\\
		T(\v{r})^\dagger & -i v_F \v{\sigma} \cdot \v{\nabla}
	\end{pmatrix}_{UL}, \\
	\quad T(\v{r}) & =
	w_{AB}
	\begin{bmatrix} 
		\kappa U_0(\v{r}) & U_1(\v{r})\\
		\overline{U_1(-\v{r})} & \kappa U_0(\v{r})
	\end{bmatrix}_{AB},
 \end{aligned}
	\label{eq:TBG_hamiltonian_appendix}
\end{equation}
where ``UL" and ``AB" specify layer and sublattice blocks respectively. We take $w_{AB} = \SI{110}{meV}$, and $0 \le \kappa \le 1$ is the chiral parameter. 

Here, $-i v_F \v{\sigma} \cdot \v{\nabla}$ is the standard continuum model of graphene for the top and bottom layers. Rotations of the Pauli matrices are neglected, $\bsigma_{\pm \theta/2} \to \bsigma$,  due to their negligable effect at small twist angles. Though for the $\kappa = 0$ chiral model of interest they may also be removed by a unitary transformation \cite{tarnopolskyOriginMagicAngles2019a}. 

The functions
\begin{equation}
\begin{aligned}
	U_0(\v{r}) & = e^{-i \bq_1 \cdot \br}+ e^{-i \bq_2 \cdot \v{r}} + e^{-i\bq_3 \cdot \v{r}},
 \\
	U_1(\v{r}) & = e^{-i \bq_1 \cdot \br}+ e^{-2\pi i/3} e^{-i \bq_2 \cdot \v{r}} + e^{-4\pi i/3} e^{-i\bq_3 \cdot \v{r}}
 \end{aligned}
\end{equation}
encode the tunneling between layers.  The tunneling wavevectors, which connect the Dirac points in each layer, are $\bq_1 = k_\theta(0,-1)$ and $\bq_{2,3} = k_\theta \left(\pm \frac{\sqrt{3}}{2}, \frac{1}{2}\right)$, where $k_\theta = \frac{4 \pi}{3 L_M}$ and $\v g_{1,2} = \bq_{2,3} - \bq_1$.

We now consider the effect of strain. Experimental samples of TBG possess both external and internal strain. Externally imposed strain (e.g. from the substrate) distorts the moir\'e unit cell itself, and can significantly shift the energetics and ground state phases of the system \cite{bi_designing_2019,parkerStraininducedQuantumPhase2021a}. Internal strain, by contrast, leaves the moir\'e unit cell unchanged and expands the AB regions while shrinking the AA regions. As internal strain is caused by the action of one graphene sheet on the other, it is necessarily present even in pristine and completely isolated samples. Physically, this is because Bernal (AB) stacking is energetically preferable to AA stacking. Therefore TBG locally displaces towards Bernal stacking everywhere except the centers of the AA regions, where the lattice instead buckles out of plane. This effect has been carefully studied in \textit{ab initio} calculations and observed in experiments. As both AB regions growing and AA regions buckling serve to reduce the AA tunnelling, it is often modelled phenomenologically by reducing the chiral ratio from $\kappa = 1$ to $\kappa = 0.5-0.8$.

Here we take a more microscopic perspective. We consider a moir\'e-periodic displacement field \cite{namLatticeRelaxationEnergy2017}
\begin{equation}
	\v{u}(\v{r}) = \v{u}(\v{r})^+ - \v{u}(\v{r})^-  =\sum_{\v{g}} \v{u}_{\v{g}} e^{i\v{g}\cdot\v{r}}
\end{equation}
which is equal and opposite between the two layers $\pm$. We take parameters $\v{u}_{\v{g}}$ from Ref. \cite{namLatticeRelaxationEnergy2017} (see also \cite{carrMinimalModelLowenergy2019}). This, in turn, induces a moir\'e-periodic pseudo-vector potential 
\begin{equation}
	\mathcal{A}^{\pm}(\v{r}) = \frac{\beta}{2a} (u_{xx}^{\pm}-  u_{yy}^{\pm}, -2u_{xy}^{\pm}), \quad \v{u}_{\mu\nu}(\v{r})^\pm = \frac{1}{2} [ \partial_{r_\mu} u_\nu^{\pm} + \partial_{r_\nu} u_{\mu}^\pm],
\end{equation}
and psuedo-magnetic field $\mathcal{B} = d \mathcal{A}$, shown in Fig. \ref{fig:strained_TBG}. We note this microscopically realistic psuedomagnetic field averages to zero, but has peaks of $\sim \SI{80}{\tesla}$ \cite{namLatticeRelaxationEnergy2017} --- a significant perturbation.

The Hamiltonian for (internally) strained TBG is therefore
\begin{equation}
	\hat{H}(\v{r})  = \hat{H}_{\mathrm{TBG}}(\v{r}) + v_F
\begin{pmatrix} 
	\sigma^\mu \mathcal{A}_\mu^{+}(\v{r}) & 0\\
	0 & \sigma^\mu \mathcal{A}_\mu^{-}(\v{r})
\end{pmatrix}.
\end{equation}
Permuting the layer and sublattice indices, we can write this as
\begin{equation}
\begin{aligned}
    \hat{H}(\v{r}) & = \begin{pmatrix}
        \kappa T_0(\v{r}) & D_{\mathrm{TBG}}^\dagger\\
        D_{\mathrm{TBG}} & \kappa T_0(\v{r})
    \end{pmatrix}, \\ 
    T_0(\v{r}) & = \begin{bmatrix}
        0 & U_0(\v{r})\\
        \overline{U_0(-\v{r})}
    \end{bmatrix},\\
     D_{\mathrm{TBG}} & = 
     \begin{bmatrix}
      2 \chi_+^\mu(\br)( -i\partial_\mu + \mathcal{A}_\mu^+) & \alpha U_1(\v{r})\\
      \alpha U_1(-\v{r}) & 2 \chi_-^\mu(\br) ( -i \partial_\mu + \mathcal{A}_\mu^-)
      \end{bmatrix},
      \end{aligned}
      \label{eq:internally_strained_TBG_full}
\end{equation}
where $\chi_{\pm}^\mu(\v{r}) = \tfrac{1}{2}[e_{1 \pm}^\mu + i e_{2 \pm}^\mu]$. Replacing $\mathcal{A}(\v{r}) \to \alpha_s \mathcal{A}(\v{r})$ and taking $\kappa \to 0$ gives Eq. \eqref{eq:strained_chiral_limit} in the main text.

Eq. \eqref{eq:internally_strained_TBG_full} retains the symmetries of unstrained TBG, including $C_{3z}$, $C_{2x}$, and particle-hole symmetries. This is because we consider strains which are equal and opposite in the two layers, and because the displacement field $\v{u}^{\pm}(\v{r})$ is itself $C_3$ and $C_{2x}$ symmetric (Fig. \ref{fig:strained_TBG}) --- and hence $\mathcal{A}(\v{r})$ is as well.

  \section{Review of traditional band geometry}\label{sec:band_geom_review}
  
  This section will review ``traditional" results on band geometry and its relationship to the FQHE. We start off by recalling the definitions of band geometry, and the key results on how to mimic a Landau level from conditions on band geometry. One of these conditions --- the determinant condition --- can be understood as a $\v{k}$-dependent generalization of the trace condition. However, we immediately argue that the determinant condition is not physically meaningful in the context of FQHE physics. Our treatment is entirely self-contained.

  For ease of referencing, we copy the band geometry definitions from the main text below.
  \begin{equation}
    \begin{aligned}
      \tilde{\eta}^{\mu \nu}(\k) & = \sum_{a}\bra{\partial^{k_\nu} u_{\bk b}} Q(\v k) \ket{\partial^{k_\mu} u_{\bk a} }, \\
      Q(\bk) & = 1-\sum_{a} \ket{u_{\bk a }} \bra{u_{\bk a}} \\
      g_\FS^{\mu \nu}(\bk) &= \Re \tilde{\eta}^{\mu \nu}(\bk), \quad \F(\bk)  \varepsilon^{\mu \nu}= - \frac{1}{2}\Im \tilde{\eta}^{\mu \nu}(\bk).
  \end{aligned}
    \label{eq:quantummetric}
  \end{equation}

\subsection{Landau Level Mimicry}

  The FS metric and Berry curvature for the LLL are given by 
  \begin{equation}
      g_{\text{FS, } 0}^{\mu \nu}(\bk) = \frac{1}{2}\ell^2 \delta^{\mu \nu}, \qquad \F_0 = \ell^2
      \label{eq:LLLgeometry}
  \end{equation}
  where $\ell$ is the magnetic length \cite{ozawaRelationsTopologyQuantum2021,ledwithStrongCouplingTheory2021}. 
Roy showed that the Girvin-Macdonald-Platzmann algebra \cite{girvinMagnetorotonTheoryCollective1986} of density operators holds if the Fubini-Study metric and the Berry curvature are the same as those of the LLL \cite{royBandGeometryFractional2014a}. If, additionally, the band is flat with Chern number $\pm 1$, then the many-body problem of electrons interacting in the LLL is fully reproduced and we expect to obtain an FCI in the band of interest. 

While a full mimicking of the LLL band geometry is required for Roy's result, partial results may or may not be sufficient. Each of the following conditions have been proposed as potential conditions, perhaps sufficient, or necessary, or both, for ideality (see e.g. \cite{jacksonGeometricStabilityTopological2015}).
\begin{enumerate}
  \item[(G1)] The Berry curvature is $\bk$-independent.
  \item[(G2)] The Fubini-Study metric is $\bk$-independent.
  \item[(G3)] The determinant inequality $\int d^2 \bk \sqrt{\det g_\FS} \geq \pi \abs{C}$ is saturated (determinant condition).
  \item[(G4)] The trace inequality $\int d^2 \bk \tr g_\FS(\bk) \geq 2\pi \abs{C}$ is saturated (trace condition).
    \label{GSH_equivstatements}
\end{enumerate}
As the geometry of the LLL \eqref{eq:LLLgeometry} obeys $\tr g_{\text{FS, } 0}(\v{k}) = \ell^2 = \F_0 $ and $\sqrt{\det g_{\text{FS, } 0}(\v{k})} = \ell^2/2$, all four conditions are exactly satisfied.
 The $\bk$-independence of $g_{FS}$ and $\mathcal{F}$ originates from continuous magnetic translation symmetry, and the relationship (G4) between the FS metric and the Berry curvature can be understood through vortexability, as discussed in the main text.

We note that there are relationships between the above conditions that we will now review. The inequality in condition (G3) comes from noting that $\tilde{\eta}$ is a positive semidefinite matrix and computing 
\begin{equation}
  \det \tilde{\eta} = \det g_\FS - \frac{1}{4}\abs{\F}^2 \geq 0.
  \label{eq:detinequality_pf1}
\end{equation}

The inequality in (G3) then follows from taking a square root, integrating over $\v{k}$, and applying the fact
\begin{equation}
  \int d^2\bk \abs{\F} \geq \int d^2 \bk \F = 2\pi C.
  \label{eq:detinequality_pf2}
\end{equation}
Therefore the inequality in (G3) is saturated if and only if $\det \tilde{\eta} = 0$ and $\F$ does not change sign. The inequality in (G4) then follows 
by applying the arithmetic mean-geometric mean (AM-GM) inequality on the real \& positive eigenvalues of $g_{FS}$, which yields $ \tr g_\FS \geq 2\sqrt{\det g_\FS } $.
 The trace condition (G4) therefore implies the determinant condition (G3) and is satisfied if and only if $g^{\mu \nu}_\FS = \pm \frac{1}{2} \F \delta^{\mu \nu}$ such that the AM-GM inequality is saturated for the eigenvalues of $g$ and $\F$ does not change sign. Additionally, this implies that (G1) and (G4) together imply (G2). Thus, if (G1) and (G4) are satisfied, all four conditions are. 

There is a rich structure associated with conditions (G3-4). In particular, the determinant condition is saturated if and only if the Berry curvature does not change sign
\footnote{Note that sign changes of the Berry curvature are somewhat singular when the determinant condition is satisfied because they come with a degenerate metric $\det g_\FS = 0$, see \eqref{eq:detinequality_pf1}. If one only demands $4\det g_\FS(\bk) = \F(\bk)^2$ for all $\bk$, rather than $\int d^2\bk \sqrt{\det g_\FS} = \pi \abs{C}$, then sign changes of the Berry curvature are allowed.}
and there is a complex vector $\omega_\mu(\bk)$ that varies smoothly with $\bk$ that satisfies the following equivalent conditions.  
\begin{enumerate}
  \item[(D1)] The vector $\omega_\mu(\bk)$ is a zero mode of the quantum metric: $\omega_\mu(\bk) \tilde{\eta}^{\mu \nu}(\bk) = 0$.
  \item[(D2)] The periodic wavefunctions satisfy $Q(\bk) \ov{\partial}_\sk \ket{u_\bk} = 0$ with $\ov{\partial}_\sk = \omega_\mu(\bk) \partial^{k_\mu}$.
  \item[(D3)] The Bloch wavefunctions satisfy $\sz_\bk \ket{\psi_{\bk a}} = \P \sz_\bk \ket{\psi_{\bk a}}$ where $\sz_\bk = \omega_\mu(\bk) r^\mu$.
    \label{detcond_equivstatements}
\end{enumerate}

The trace condition (G4) is a special case of the determinant condition with $g^{\mu \nu}_\FS \propto \delta^{\mu \nu}$ and corresponds to the $\bk$-indepdenent $\omega_\mu(\bk) = \begin{pmatrix}1 & i \end{pmatrix}^T$ such that $\sz_\bk = z = x+iy$.

The condition (D3) is a weaker version of vortexability where the vortex function is allowed to be $\bk$ dependent. To our knowledge, this weaker vortexability condition does not enable us to make any pseudopotential arguments. Furthermore, we will soon argue that ideality conditions for the FQHE should be invariant under forgetting translation symmetry, and the determinant condition is not --- unless $\omega_\mu$ is $\bk$-independent (see Sec. \ref{subsec:detcond_folding}). A generic $\bk$-independent $\omega_\mu$ is a mild generalization of the trace condition; the usual trace condition may be recovered by the linear transformation on $\bk$ that makes the Fubini-Study metric proportional to the identity matrix. 

Let us now prove the equivalence of (D1-3) to the determinant condition with single-sign Berry curvature. We first prove this for (D1) and then show (D1)$\iff$(D2), (D2)$\iff$(D3).
Note that the determinant condition is satisfied if and only if $\det \tilde{\eta}(\bk) = 0$ for all $\bk$ and the Berry curvature has the same sign throughout the BZ; see the discussion around \eqref{eq:detinequality_pf1}, \eqref{eq:detinequality_pf2}. Then the condition $\det \tilde{\eta}(\v{k}) = 0$ is satisfied if and only if there is a zero mode $\omega_\mu(\bk) \tilde{\eta}^{\mu \nu}(\bk) = 0$.

To show (D1)$\iff$(D2) we compute
\begin{equation}
\begin{aligned}
  \omega_\mu(\bk) \tilde{\eta}^{\mu \nu}(\bk) \ov{\omega_\nu(\bk)} & = \sum_a \bra{\omega_{\nu}\partial^{k_\nu} u_{\bk a}} Q(\bk) \ket{\omega_\mu \partial^{k_\mu} u_{\bk a}} \\
  & = \sum_a \norm{Q(\bk) \partial_{\sk} u_{\bk a}}^2.
  \end{aligned}
  \label{eq:vanishnorm}
\end{equation}
The vanishing of the left and right hand sides are equivalent to (D1) and (D2) respectively.

We finish by showing (D2)$\iff$(D3) using an argument that is very similar to that of the trace condition vortexability in the main text; we include the details here for completeness. We begin with the product rule
  \begin{equation}
    -i \bnabla_\bk \ket{\psi_\bk^a} = \hat{\br} \ket{\psi_\bk^a} +  e^{i \bk \br}  (-i\bnabla_\bk) \ket{u_\bk^a}.
      \label{eq:productrule}
  \end{equation}
  Acting with $\Q := 1-\P = 1-\sum_{\bk a} \ket{\psi_{\bk a}} \bra{\psi_{\bk a}}$
   annihilates the left hand side of Eq. \eqref{eq:productrule}, giving
  \begin{equation}
      \Q \br \ket{\psi_\bk^a} =   \Q  i e^{i \bk \cdot \br} \bnabla_\bk \ket{u_\bk^a}.
  \label{eq:projected_product_rule_supp}
  \end{equation}
  For clarity, the projector $\Q$ acts on Bloch periodic wavefunctions $\ket{\psi_{\v{k}}}$ and contains a sum over $\bk$, whereas $Q(\bk)$ acts on periodic wavefunctions, appears in the definition of quantum geometry \eqref{eq:quantummetric}, and contains no sum over $\bk$. Since $\bra{\psi_{\bk'}} e^{i \bk \cdot \br} \nabla_\bk \ket{u_\bk}$ vanishes for $\bk \neq \bk'$ by translation symmetry, we have $\Q e^{i \bk \cdot \br} \bnabla_\bk \ket{u_{\bk}} = e^{i \bk \cdot \br} Q(\bk) \bnabla_\bk \ket{u_{\bk}}$, a rewriting of the right hand side of \eqref{eq:projected_product_rule_supp}. We therefore have
\begin{equation}
  \Q r^\mu \ket{\psi_{\bk a}} =  e^{i \bk \cdot \br}Q(\bk) i\partial^{k_\mu} \ket{u_{\bk a}},
  \label{eq:projected_product_rule_supp_final}
\end{equation}
such that dotting each side with $\omega_\mu(\bk)$ gives
\begin{equation}
  \Q \sz_\bk \ket{\psi_{\bk a}} = 0 \iff Q(\bk) \ov{\partial}_{\sk} \ket{u_{\bk a}}
  \label{eq:detcond_vortexability_equivalence}
\end{equation}
which establishes the equivalence between (D2) and (D3).

In Ref. \cite{meraKahlerGeometryChern2021}, the condition (D2) was interpreted in the context of K\"{a}hler geometry. In particular, the periodic wavefunctions $\ket{u_{\bk a}}$ determine a map from the Brillouin Zone to complex projective space (such that the unphysical phase of the wavefunction is quotiented away). This map is a holomorphic function of the complex coordinate $\sk$, where $\sk$ solves the Beltrami equation $\ov{\partial}_\sk \sk = 0$, if and only if the condition (D2) is satisfied with a ``non-degenerate" $\omega_\mu(\bk)$. By non-degenerate we mean that the metric $g^{(\omega)}_{\mu \nu} = \Re \ov{\omega_\mu} \omega_\nu$ is non-degenerate (i.e. bounded with no zero modes). It is possible to show that this is the case if the Fubini-Study metric is bounded with no zero modes, as $g_\FS \propto (g^{(\omega)})^{-1}$, but this will not be important for us.

Through such a holomorphic map, the K\"ahler structure of complex projective space is pulled back to the Brillouin Zone and the pull backs of the Riemannian metric and symplectic form of complex projective space may be identified as the Fubini-Study metric and the Berry curvature respectively \cite{meraKahlerGeometryChern2021}.

\subsection{Holomorphic Bloch wavefunctions
}
\label{subsec:holo_Bloch}

The determinant and trace conditions enable us to perform a gauge transformation, that is in general non-unitary and non-periodic,
\begin{equation}
\ket{u_{\bk}} \to \ket{\su_\sk} = S(\bk) \ket{u_{\bk}}, 
\end{equation}
with matrix multiplication in band indices, such that the new wavefunction is a holomorphic function of the coordinate $\sk$
\cite{meraKahlerGeometryChern2021}
\begin{equation}
\ov{\partial}_\sk \ket{\su_\sk} = \omega_\mu(\bk) \partial^{k_\mu} \ket{\su_\sk} = 0.
\label{eq:hologauge_supp}
\end{equation}  
In this section we will explicitly construct such an $S(\bk)$. 

While $S(\bk)$ is in general non-unitary and non-periodic, this is a powerful result that we will use to prove a uniqueness result on vortex functions (Prop. \ref{prop:uniqueness} below). Indeed, as discussed in the main text, there is always a coordinate frame for which vortexable bands satisfy (D1) and (D2) with a $\bk$-independent $\omega_\mu$ (see also Prop \ref{prop:holo_gauge_choice}). We imagine $\sk$-holomorphicity will be a future calculational tool as well for working with such bands, so we elaborate on how to choose such a gauge here. 

To obtain \eqref{eq:hologauge_supp} we will combine (D2), $Q(\bk) \ov{\partial}_\sk \ket{u_\bk} = 0$, a gauge invariant statement, with a choice of $S(\bk)$ such that
\begin{equation}
\begin{aligned}
    \ov{\partial}_\sk \underbrace{(S_\bk \ket{u_\bk})}_{\ket{\su_\sk}} 
    & = [Q(\bk) + P(\bk)] \ov{\partial}_\sk (S_\bk \ket{u_\bk}) \\
    & = 
    P(\bk) \ov{\partial}_\sk (S_\bk \ket{u_\bk}) = 0, \\
    & \iff (\ov{\partial}_\sk + i \ov{A}_\sk) S_\bk = 0,
    \end{aligned}
    \label{eq:eqn_for_Sk_supp}
\end{equation}
where $\ov{A}_\sk(\bk) = \omega_\mu A^\mu(\bk)$ is a matrix-valued function of $\bk$, and $A^\mu_{ab}(\bk) = -i \bra{u_{\bk b}} \partial^{k_\mu} \ket{u_{\bk a}}$ is the Berry connection. We assume the Berry connection is smooth in $\bk$ (and therefore not periodic under translations by reciprocal lattice vectors). Such a gauge choice is always possible. 

We will solve the right side of \eqref{eq:eqn_for_Sk_supp} with an explicit ansatz for $S_\bk$. We begin by noting the identity
\begin{equation}
\begin{aligned}
    \ov{\partial}_\sk W[f](\bk) & = f(\bk), \\
    W[f](\bk) & = \frac{1}{2\pi i} \int_{\mathbb{C}} \frac{f(\sk - \zeta)}{\zeta} d\ov{\zeta} d\zeta.
    \end{aligned}
    \label{eq:inverseofdelbar}
\end{equation}
Note that $f$ does not need to be holomorphic.
For completeness we pause to prove \eqref{eq:inverseofdelbar}. We excise an arbitrarily small disk $D_\varepsilon$ centered at the origin with radius $\varepsilon$ (where the denominator becomes singular) and compute, using Stokes' theorem
\begin{equation}
\begin{aligned}
    \ov{\partial}_{\sk} W[f](\sk) & = \lim_{\varepsilon \to 0}
    \frac{1}{2\pi i} \int_{\mathbb{C}-D_{\varepsilon}} \ov{\partial}_\sk\frac{f(\sk - \zeta)}{\zeta} d \ov{\zeta} \wedge d{\zeta} \\
    & = -\lim_{\varepsilon \to 0} \frac{1}{2\pi i} \int_{\mathbb{C}-D_{\varepsilon}} \frac{\ov{\partial}_\zeta f(\sk - \zeta)}{\zeta} d \ov{\zeta} \wedge d{\zeta} \\
    & = \lim_{\varepsilon \to 0} \frac{1}{2\pi i} \oint_{\partial D_{\varepsilon}}\frac{f(\sk - \zeta)}{\zeta} d{\zeta} \\
    & = f(\sk) \lim_{\varepsilon \to 0 } \frac{1}{2\pi i } \oint_{D_\varepsilon} \frac{d \zeta}{\zeta}  \\
    & = f(\sk).
   \end{aligned} 
   \label{eq:holo_stokes_calculation}
\end{equation}
To apply Stokes' theorem, we used the differential-form calculation $d(\frac{f}{\zeta} d \zeta) = \frac{\partial_{\ov \zeta} f}{\zeta} d \ov{\zeta} \wedge d \zeta$ away from $\zeta = 0$. 
The negative sign in the second line originates from the chain rule and disappears in the third line due to change in orientation from $\partial(\mathbb{C} - D_\varepsilon)$ to $\partial{D_\varepsilon}$. Finally, we pulled $f(\sk - \zeta) \approx f(\sk)$ out of the integral using the $\varepsilon \to 0$ limit in which the integral domain involves arbitrarily small $\zeta$.

We now write the ansatz
\begin{equation}
    S_\bk = 1 - i W[h](\bk)
\end{equation}
under which we may solve \eqref{eq:eqn_for_Sk_supp} for $h$ iteratively:
\begin{equation}
\begin{aligned}
    & h  = \ov{A}_\sk - i \ov{A}_\sk W[h], \\
    & = \ov{A}_\sk - i \ov{A}_\sk W[\ov{A}_\sk] + (-i)^2 \ov{A}_\sk W[\ov{A}_\sk W[\ov{A}_\sk]] + \ldots.
    \end{aligned}
    \label{eq:iterativesolution}
\end{equation}
We may view the above solution as a generalization of the path ordered exponential used to solve real first order matrix differential equations. To show convergence of Eq. \eqref{eq:iterativesolution}, we first show that it converges locally and then patch together local solutions. We may partition the total space $\mathbb{C}$ (that $\sk$ lives on) with an open cover of sets $U_i$ for $U_i$ arbitrarily small. Note that on a set $U_i$ of diameter $\delta$ we have $\abs{W[f]} \leq \delta \max_{U_i} \abs{f}$, such that \eqref{eq:iterativesolution} converges for sufficiently small $\delta$. It remains to patch the solutions on the various $U_i$ together, but there is no obstruction to this because $\mathbb{C}$ is contractible. Explicitly, on intersections $U_i \cap U_j$, the solutions differ by a holomorphic matrix: $S_i(\bk) = F_{ij}(\sk) S_j(\bk)$. To see this, apply $\ov{\partial}_\sk$ to $F_{ij} = S_{j}^{-1} S_i$ and use \eqref{eq:eqn_for_Sk_supp}. The collection of matrices $F_{ij}(\sk)$ may be interpreted as transition functions for a holomorphic vector bundle over $\mathbb{C}$. But since $\mathbb{C}$ is contractible, there exists trivializing functions $g_i(\sk)$ such that $F_{ij}(\sk) = \ov{g_i(\sk)} g_j(\sk)$. Then the matrix $S_\bk = g_i(\sk) S_i(\bk)$ is independent of the patch $i$, and is thus globally defined.

We have now shown the existence of a complexified gauge $S_{\bk}$ so that $\ket{\su_\sk} = S_\bk \ket{u_\bk}$ is $\sk$-holomorphic.

\subsection{The determinant condition is not folding invariant}
\label{subsec:detcond_folding}

One filtration technique on natural ideal conditions for fractional quantum Hall states is invariance under the process of ``forgetting" translation symmetry. Indeed, opting to use the maximally sized Brillouin zone is a choice; the fractional quantum Hall effect is just as likely to occur if we use a different description to describe the single particle wavefunctions. 

Another way to motivate this requirement is that the FQHE is stable under weak perturbations, and so the single particle band-geometric heuristics should not drastically change under such perturbations, for example an arbitrarily weak perturbation that breaks translation symmetry to a subgroup and folds the Brillouin Zone onto itself.

 Consider enlarging the real space unit cell area by a factor of $M$. This increases the number of bands from $N$ to $MN$ and causes $M$ vectors $\bQ_m$ in the original first Brillouin Zone to become reciprocal lattice vectors, such that the new first Brillouin Zone is $M$ times smaller. This ``folding" of the Brillouin Zone corresponds to a new labeling of the single particle wavefunctions
\begin{equation}
    \ket{\psi^f_{\bk (a,m)}} = \ket{\psi_{\bk + m \bQ_m, a}}.
\end{equation}
for $\bk$ in the folded Brillouin zone, which is $M$ times smaller than the original Brillouin Zone. Correspondingly
\begin{equation}
    \ket{u^f_{\bk (a,m)}} = e^{i \bQ_m \cdot \br}\ket{u_{\bk + m \bQ_m, a}}.
\end{equation}
The ``forgotten" translation symmetry implies that $\braket{\tilde{u}_{\bk(a,m)}|\tilde{u}_{\bk' (a',m')}} = 0$ for $m \neq m'$, even if $\bk \neq \bk'$. The new quantum metric can therefore be computed as
\begin{equation}
    \tilde{\eta}_f^{\mu \nu}(\bk) = \sum_m \tilde{\eta}^{\mu \nu}(\bk + \bQ_m)
\end{equation}
such that it is now periodic in the new first Brillouin Zone.

We will now show that the determinant condition is not invariant under the operation of forgetting translation symmetry and folding the Brillouin Zone, except when $\omega_\mu(\bk)$ is $\bk$-independent. In this case the trace condition in linearly transformed coordinates is satisfied. We use condition (D1): $\omega_\mu(\bk) \tilde{\eta}^{\mu \nu}(\bk) = 0$. For this to be true for $\tilde{\eta}^f(\v{k})$, which is a sum of positive definite matrices $\tilde{\eta}(\bk + \bQ_m)$, we must have that the zero modes $\omega(\bk + \bQ_m)$ all coincide. If we consider a thermodynamically large system and forget all translation symmetry, then the vectors $\bQ_m$ form a dense subset of the Brillouin Zone such that we must have a $\bk$-independent $\omega_\mu$ as claimed.

Note that our result --- that the determinant condition is folding invariant if and only if $\omega_\mu(\bk)$ is constant --- was first obtained by Refs. \cite{meraKahlerGeometryChern2021,ozawaRelationsTopologyQuantum2021} from a mathematical perspective of satisfying the determinant condition on the twist angle torus.  Their twist angle torus effectively emerges when we forget all translation symmetry, such that the entire system is a single unit cell and the Bloch boundary conditions and the twist angle boundary conditions may be identified.

\section{Uniqueness of the vortex function}
\label{sec:vortex_uniqueness}

This section is devoted to proving a uniqueness result on vortexable bands: Prop. \ref{prop:uniqueness}. We show that under mild assumptions --- a `generic' continuum model with discrete translation symmetry  --- vortex functions are unique.  
 Some restriction on the electron density is necessary to avoid pathological cases: the vortex function may take arbitrary values where the band has no density. For example, in the strict tight binding limit only the values of the vortex function at atomic sites matter. We will take the electron density to be finite continuous and almost never zero, which is sufficient for our proof and is the typical situation in continuum models. The assumption of lattice translation symmetry can perhaps be weakened, though we emphasize that the unit cell of the lattice can be arbitrarily large and we do not restrict the number of Bloch bands. We expect most condensed matter systems to be translation symmetric in the limit where the unit cell is taken to be thermodynamically large. 

We will have to prove some prerequisite results before directly addressing uniqueness. We start with a simple result on the form of vortexable bands with translation symmetry.

 \prop{Consider a vortexable band with discrete translation symmetry: $\br \to \br + \ba$ and $\partial_\mu \sz(\br + \ba) = \partial_\mu \sz(\br)$. Then $\sz(\br + \ba) = \sz(\br) + \omega_{\mu} a^\mu$ for some complex vector $\omega_{\mu}$.
\label{prop:general_translation} 
 }
 \proof{From $\partial_\mu \sz(\br + \ba) = \partial_\mu \sz(\br)$ we conclude that $\sz(\br + \ba) = \sz(\br) + \beta_\ba$ for an $\ba$ dependent constant $\beta_\ba$. Iterating translations, we find $\beta_{\ba_1 + \ba_2} = \beta_{\ba_1} + \beta_{\ba_2}$ such that $\beta$ is a linear function on the lattice. We may therefore write $\beta_\ba = \omega_{\mu} a^\mu$ for some complex-valued vector $\omega_\mu$. \qed }
 
 \bigskip

We will find it convenient to use a unit-cell averaged version of the vortex geometry to characterize $\omega_\mu$.

\lemma{Consider a vortexable band with discrete translation symmetry $\sz(\br + \ba) = \sz(\br) + \omega_\mu a^\mu$ (Prop. \ref{prop:general_translation}). We have
\begin{equation}
    \int_{\uc} \Omega(\br) d^2 \br= \Omega^{(\omega)} A_{\uc} = -i\varepsilon^{\mu \nu} \ov{\omega_\mu} \omega_\nu A_{\uc}.
\end{equation}
Here $\Omega = -i\varepsilon^{\mu \nu} \partial_\mu \ov{\sz} \partial_\nu \sz$ is the vortex chirality, the integral is over the unit cell, $A_\uc$ is the unit cell area, and $\Omega^{(\omega)} = -i\varepsilon^{\mu \nu} \ov{\omega}_\mu \omega_\nu$. Furthermore, the metric $g^{(\omega)}_{\mu \nu} = \Re \ov{\omega}_\mu \omega_\nu$ is non-degenerate ($\det g^{(\omega)} \neq 0)$.
\label{lem:vortex_stokes}
}
\proof{
First we write $\Omega$ as a curl
 \begin{equation}
 \begin{aligned}
    \Omega & = -i\varepsilon^{\mu \nu} \partial_\mu \ov{\sz} \partial_{\nu} \sz = -i\varepsilon^{\mu \nu} \partial_\mu( \ov{\sz} \partial_\nu \sz ) = \hat{\v z} \cdot \bnabla \times \bA, \\ 
    A_\mu & = -i\ov{\sz} \partial_\mu \sz,
    \end{aligned}
 \end{equation} 
 such that we may apply Stokes' theorem
 \begin{equation}
     \int_{\uc} \Omega d^2 \br = \oint_{\partial \uc} \bA \cdot d\br,
 \end{equation}
 where the line integral is taken counter-clockwise around the parallelogram defined by the points $\br, \br + \ba_1, \br + \ba_1 + \ba_2, \br + \ba_2$. We may group the paths associated with the opposite sides of the parallelogram and use the boundary condition $\ov{\sz}(\br + \ba) = \ov{\sz}(\br) + \ov{\omega}_\mu a^\mu$. We then obtain 
 \begin{equation}
 \begin{aligned}
     \oint_{\partial \uc} \bA \cdot d \br 
     & = -i\int_{\br + \ba_1}^{\br + \ba_1 + \ba_2} \ov{\omega}_\mu a_1^\mu \bnabla \sz \cdot d\br \\
     & \quad -i \int_{\br + \ba_1 + \ba_2}^{\br + \ba_2} \ov{\omega}_\nu a_{2}^\nu \bnabla \sz \cdot d\br  \\
     & = -i\ov{\omega}_\mu a_1^\mu \omega_\nu a_2^\nu - \ov{\omega}_\mu a_2^\mu \omega_\nu a_1^\nu \\
     & = -i(\ov{\omega}_\mu \omega_\nu - \ov{\omega}_\nu \omega_\mu) a_1^\mu a_2^\nu \\
     & = -i\varepsilon^{\rho \sigma} \ov{\omega}_\rho \omega_\sigma A_{\uc},
     \end{aligned}
 \end{equation}
 where we used that two-dimensional antisymmetric matrices $B_{\mu \nu}$ are proportional to $\varepsilon_{\mu \nu}$ and therefore satisfy $B_{\mu \nu} = \frac{1}{2} \varepsilon^{\rho \sigma} B_{\rho \sigma} \varepsilon_{\mu \nu}$. We also identified $\varepsilon_{\mu \nu} a_1^\mu a_2^\nu = A_{\uc}$.

For the last part of the proposition we use the rank one Hermitian matrix $\eta^{(\omega)}_{\mu \nu} = \ov{\omega}_\mu \omega_\nu$. Computing $0 = \det \eta^{(\omega)} = \det g^{(\omega)} - \abs{\Omega^{(\omega)}}^2$ then implies that $\det g^{(\omega)} \neq 0$.
 
 \qed
}
\bigskip

\prop{
Suppose we have a vortexable band with discrete translation symmetry $\br \to \br+\ba$, where the vortex function satisfies $\sz(\br + \ba) = \sz(\br) + \omega_\mu a^\mu$ (Prop \ref{prop:general_translation}). Define periodic wavefunctions
\begin{equation}
    \ket{u'_{\bk a}} = e^{-i \bk \cdot \br'} \ket{\psi_{\bk a}},
\end{equation}
using a periodic diffeomorphism $\br'(\br + \ba) = \br'(\br) + \ba$. We may choose $\br'(\br)$ such that
\begin{enumerate}
    \item $Q(\bk) \ov{\partial}_\sk \ket{u_{\bk a}} = 0$ where $\ov{\partial}_\sk = \omega_\mu \partial^{k_\mu}$
    \item There exists a gauge choice $\ket{\su_\sk} = S_\bk \ket{u_\bk}$, where $S_{\bk}$ is invertible but generically non-unitary, so that
    \begin{equation}
        \ov{\partial}_{\sk} \ket{\su_\sk} = 0,
    \end{equation}
    i.e. the periodic wavefunctions are holomorphic functions of $\sk$.
\end{enumerate}
Here $\sk$ is a linear function of $\v{k}$ such that $\ov{\partial}_\sk \sk = 0$.
\label{prop:holo_gauge_choice}
}

\proof{
The coordinate system $\br'(\br)$ is obtained by first choosing ``non-periodic" coordinates $\br_{\sz}(\br)$ where $\sz = x_\sz + i y_\sz$ using the fact that $\sz: \mathbb{R}^2 \to \mathbb{C}$ is a diffeomorphism [see below Eq. \eqref{eq:non-degeneracy_condition}]. These coordinates satisfy
\begin{equation}
    r_\sz^\mu(\br + \ba) = r^\mu_\sz(\br) + J^\mu_\nu a^\nu
\end{equation}
where $J^x_\mu = \Re \omega_\mu$ and $J^y_\mu = \Im \omega_\mu$. We have $J^T J = g^{(\omega)}$, and $\det g^{(\omega)} \neq 0$ (Lemma \ref{lem:vortex_stokes}), so $J$ is invertible. We then define $(r')^\mu = (J^{-1})^\mu_\nu r_\sz^\nu$, whereupon
\begin{equation}
    \br'(\br + \ba) = \br'(\br) + \ba
\end{equation}
as claimed. 

With this choice $\sz = \omega_\mu r^{\prime \mu}$. Then the equivalence (D1)$\iff$(D2)$\iff$(D3) in Sec. \ref{sec:band_geom_review}, applied to the wavefunctions $\ket{u'_{\bk a}} = e^{-i \bk \cdot \br'}\ket{\psi_{\bk a}}$, implies that $\omega_\mu$ is a left zero mode of the quantum metric and $Q(\bk) \ov{\partial}_\sk \ket{u_{\bk a}} = 0$. Next, the non-degeneracy $\det g^{(\omega)} \neq 0$ together with the discussion of Sec. \ref{subsec:holo_Bloch} enables us to construct an $S_\bk$ such that $\ov{\partial}_\sk \ket{\su_\sk}$ = 0. \qed 
}
\bigskip 

A canonical choice of holomorphic coordinate $\sk$ is
\begin{equation}
    \sk = i \left(\Omega^{(\omega)}\right)^{-1}\varepsilon^{\mu \nu}\omega_\mu k_\nu,
    \label{eq:holomorphic_k}
\end{equation}
where $\Omega^{(\omega)} = -i \varepsilon^{\mu \nu} \ov{\omega}_\mu \omega_\nu$ from Lemma \ref{lem:vortex_stokes}. The definition \eqref{eq:holomorphic_k} was chosen so that
\begin{equation}
   \ov{\partial}_\sk \sk = \omega_\mu \partial^{k_\mu} \sk = 0, \qquad \partial_{\sk} \sk = \ov{\omega}_\mu \partial^{k_\mu} \sk = 1.
   \label{eq:standard_holo_derivatives}
\end{equation}

A conceptual point worth emphasizing is that vortexability $\sz \ket{\psi} = \P \sz \ket{\psi}$ remarkably does not depend on the Hilbert space inner product $\langle \cdot | \cdot \rangle$ --- despite its dependence on the orthogonal projector $\P = \P(\langle \cdot | \cdot \rangle)$. Indeed, the existence of a gauge where $\partial_\sk \su_\sk = 0$ is satisfied is completely independent of the Hilbert space inner product, its equivalence to vortexability notwithstanding.

We need one last technical Lemma before the uniqueness result. Namely, we need to understand some properties of the $\bk$-space zeros of wavefunctions in the holomorphic gauge. The existence of such zeros and their relation to the Chern number was shown by Ref. \cite{wangExactLandauLevel2021a} for a single Bloch band. Since vortexable ``bands" can contain many Bloch bands, we must prove this result for multiple bands. This follows by passing to wavefunctions
\begin{equation}
\sU_\sk(\br_1 \ldots \br_N) = \varepsilon^{a_1, \ldots a_N} \su_{\sk a_1}(\br_1) \ldots \su_{\sk a_N}(\br_N).
\label{eq:determinant_bundle}
\end{equation}
in the ``determinant-line-bundle" associated to the vector bundle of Bloch wavefunctions $\su_{\bk a}$.

\lemma{Consider a vortexable band with discrete translation symmetry $\sz(\br + \ba) = \sz(\br) + \omega_\mu a^\mu$ (Prop. \ref{prop:general_translation}), such that there are holomorphic periodic wavefunctions $\ket{\su_\sk}$ (Prop. \ref{prop:holo_gauge_choice}) where $\sk$ is given by \eqref{eq:holomorphic_k}. Then the Chern number is given by the contour integral around the boundary of the Brillouin Zone
\begin{equation}
    C = \frac{1}{2\pi i} \oint_{\partial \bz} \partial_\sk \ln  \sU_\sk(\br_1 \ldots \br_N) d \sk = N_z
    \label{eq:Chern_number_det_bundle}
\end{equation}
where  $\partial_{\sk}$ and $\sk$ are given in \eqref{eq:holomorphic_k}, \eqref{eq:standard_holo_derivatives} such that $\partial_\sk \sk = 1$. The integral counts the number of $\bk$-space zeros $N_z$, multiplicities included, of $\sU_\sk$ defined in \eqref{eq:determinant_bundle}.
\label{lem:ChernNumberZeros}}

\proof{We first show that the integral counts the number of zeros. 
Note that the logarithm $\log \sU_{\sk}$, as well as $\sU_{\sk}$, are $\sk$-holomorphic functions where they are defined (away from zeros of the logarithm). Thus we may use Cauchy's integral theorem to deform the path of the integral as we please, as long as we don't cross any zeros. We deform the path such that it makes an arbitrarily small circle $C_\alpha$ around each zero and retraces its path between zeros, such that we obtain a sum of contour integrals around each zero. Near a zero at $\sk_\alpha$ with multiplicity $p_\alpha$ we have that $U_\sk(\br_1, \ldots \br_N) \propto (\sk - \sk_\alpha)^{p_\alpha}$, which implies
\begin{equation}
\begin{aligned}
    & \frac{1}{2\pi i} \oint_{\partial \bz} \partial_\sk \ln  \sU_\sk(\br_1 \ldots \br_N) d \sk \\
    & = \sum_\alpha \frac{1}{2\pi i}\oint_{C_\alpha} p_\alpha \partial_\sk \ln (\sk - \sk_{\alpha}) \\
    & = \sum_\alpha p_\alpha = N_z.
    \end{aligned}
    \label{eq:count_zeros}
\end{equation}

We now compute \eqref{eq:count_zeros} directly on the boundary of the Brillouin Zone using $\bk$-space boundary conditions to relate it to the Chern number of the band. Specifically, we have that $\su_{\sk + \sG} $ is related by a gauge transformation $\Xi_\bG(\sk)$, a matrix in band space, to $e^{-i \bG \cdot \br}\su_{\sk}$:
\begin{equation}
  \su_{\sk + \sG} = \Xi_{\bG}(\sk)e^{-i \bG \cdot \br}\su_{\sk}.
  \label{eq:KspaceBCs}
\end{equation}
For the function \eqref{eq:determinant_bundle} we have
\begin{equation}
   \sU_{\sk + \sG}(\{\br_a\}) = \det \Xi_\bG(\sk) e^{-i \bG \cdot \sum_a \br_a} \sU_{\sk }(\{\br_a\}).
   \label{eq:KspaceBCs_bigU}
\end{equation}
We may then directly compute the contour integrate around the parallelogram $\bk$, $\bk + \bG_1$, $\bk + \bG_1 + \bG_2$, and $\bk + \bG_2$  as
\begin{equation}
\begin{aligned}
    2\pi i N_z & = \oint_{\partial \bz} \partial_\sk \ln  \sU_\sk(\br_1 \ldots \br_N) d \sk \\
    & =  \int_{\sk+\sG_1}^{\sk + \sG_1 + \sG_2} \partial_{\sk} \ln \det  \Xi_{\bG_1}(\sk - \sG_1)  \\
    & \quad + \int_{\sk + \sG_1 + \sG_2}^{\sk + \sG_2} \partial_\sk \ln \det \Xi_{\bG_2}(\sk - \sG_2) d\sk \\
    & = \ln \det \Xi_{\bG_1}(\sk + \sG_2) - \ln \det \Xi_{\bG_1}(\sk)  \\
    & \quad + \ln \det \Xi_{\bG_2}(\sk) - \ln \det \Xi_{\bG_2}(\sk + \sG_1).
    \end{aligned}
    \label{eq:number_zeros_from_bcs}
\end{equation}
Note that the dependence of the boundary conditions \eqref{eq:KspaceBCs_bigU} on $\br_a$ has dropped out due to the derivative $\ov{\partial}_\sk$.

We now compute the Chern number of the vortexable band using the boundary conditions \eqref{eq:KspaceBCs}. First we deal with the fact that our choice of holomorphic gauge has resulted in non-orthonormal band basis. The usual definition of the Berry connection $\bA_{ab} = -i \bra{u_{\bk b}} \bnabla_\bk \ket{u_{\bk a}}$ is not valid in this setting: it is only gauge covariant under \emph{unitary} gauge transformations. There is a simple fix, however, through the use of the Gram matrix $X_{ab}(\bk) = \bra{u_{\bk b}}\ket{u_{\bk a}}$ which transforms as $X(\bk) \to S_\bk X(\bk) S_\bk^\dag$ for general invertible $S_\bk$ The Berry connection
\begin{equation}
   \bA_{ab}(\bk) = -i (\bra{u_{\bk}} X^{-1})_b \bnabla_\bk \ket{u_{\bk a}},
\end{equation}
transforms as
\begin{equation}
    \bA(\bk) \to S^{-1}_\bk \bA(\bk) S_\bk -i S^{-1}_\bk \bnabla_\bk S_\bk
    \label{eq:general_gauge_covariant_transf}
\end{equation}
under a general invertible gauge transformation $S_\bk$. 
Thus, under $\bk \to \bk + \bG$ we have, using the holomorphic gauge $\ket{\su_\sk}$ and the associated boundary conditions \eqref{eq:KspaceBCs},
\begin{equation}
    \bA(\bk + \bG) = \Xi^{-1}_{\bG}(\sk) \bA(\bk) \Xi_\bG(\sk) - i \Xi^{-1}_{\bG}(\sk) \bnabla_\bk \Xi_\bG(\sk)
\end{equation}
such that
\begin{equation}
\begin{aligned}
    \tr(\bA(\bk + \bG) - \bA(\bk)) & = -i \tr \Xi^{-1}_{\bG}(\sk) \bnabla_\bk \Xi_\bG(\sk)  \\
    & = -i\bnabla_\bk \tr \log   \Xi_\bG(\sk) \\
    & = -i \bnabla_\bk \log \det \Xi_\bG(\sk).
    \end{aligned}
\end{equation}

We now compute the Chern number, which in our smooth non-periodic gauge can be obtained via Stokes' theorem
\begin{equation}
\begin{aligned}
    2\pi i C & = i\int_{\bz} d^2 \bk \F(\bk) \\
    & = i\oint_{\partial \bz} \bA \cdot d\bk \\
    & = i\int_{\bk + \bG_1}^{\bk + \bG_1 + \bG_2} (\bA(\bk) - \bA(\bk - \bG_1))\cdot d\bk\\
    & \quad +i \int_{\bk + \bG_1 + \bG_2}^{\bk + \bG_2} (\bA(\bk) - \bA(\bk - \bG_2) \cdot d \bk \\
    & = \int_{\bk + \bG_1}^{\bk + \bG_1 + \bG_2} \bnabla_\bk \ln \det \Xi_{\bG_1}(\sk - \sG_1)\cdot d\bk \\
    & \quad + \int_{\bk + \bG_1 + \bG_2}^{\bk + \bG_2} \bnabla_\bk \ln \det \Xi_{\bG_2}(\sk - \sG_2) \cdot d \bk \\
    & = \ln \det \Xi_{\bG_1}(\sk + \sG_2) - \ln \det \Xi_{\bG_1}(\sk) \\
    & \quad + \ln \det \Xi_{\bG_2}(\sk) - \ln \det \Xi_{\bG_2}(\sk + \sG_1).
    \end{aligned}
    \label{eq:Chern_number_boundaryconditions}
\end{equation}
Comparing \eqref{eq:Chern_number_boundaryconditions} and \eqref{eq:number_zeros_from_bcs} yields our desired result. \qed
}

\bigskip

We are now in a position to state and prove our uniqueness result.

\prop{A vortexable band with discrete translation symmetry $\br \to \br + \ba$, $\partial_\mu \sz = \partial_\mu \sz(\br +\ba)$ and an electron density $\rho(\br) = \sum_{\bk a} \abs{\psi_{\bk a}(\br)}^2$ that is finite, continuous, and almost never zero, has a unique vortex function up to affine transformations $\sz \to \alpha \sz + \beta$.
\label{prop:uniqueness}}
\proof{
In this proof we will restrict to $C, \Omega > 0$ without loss of generality (note that for $C=0$ the assumption that the electron density is finite and continuous is false: see Sec. \ref{subsec:chern_zero}). 

Let us assume that there are two vortex functions $\sz_1$ and $\sz_2$ (which we will eventually show must be related by an affine transformation) that satisfy
\begin{equation}
    \sz_1(\br + \ba) = \sz_1(\br) + \omega^{(1)}_\mu a^\mu,
    \quad 
    \sz_2(\br + \ba) = \sz_2(\br) + \omega^{(2)}_\mu a^\mu.
\end{equation}
By Prop \ref{prop:holo_gauge_choice}, we have periodic wavefunctions 
\begin{equation}
  \ket{\su^{(i)}_{\sk^{(i)}}} = S_{\bk}^{(i)}e^{-i \bk \cdot \br_i'} \ket{\psi_\bk},
\end{equation}
for $i = 1,2$ such that $\ket{\su^{(i)}_{\sk^{(i)}}} $ is a holomorphic function of $\sk^{(i)} = i\left(\Omega^{(\omega_i)}\right)^{-1}\varepsilon^{\mu \nu} \omega_\mu^{(i)} k_\nu$ and $S_\bk^{(i)}$ is a matrix in band-space that implements the choice of a holomorphic gauge.

The periodic wavefunctions are related as
\begin{equation}
    \ket{\su^{(1)}_{\sk_1}} = e^{-i \bk \cdot (\br_1'- \br_2')} \tilde{S}_{\bk} \ket{\su^{(2)}_{\sk_2}}.
    \label{eq:two_u_relation}
\end{equation}
where $\tilde{S}_\bk = (S^{(1)}_\bk)^{-1} S^{(2)}_\bk$. 

We wish to show that $\omega_\mu^{(1)} \propto \omega_\mu^{(2)}$. The intuition is as follows: the shape of the $\bk$-space zeros of $\su^{(1,2)}$ should be similar since they are related linearly as \eqref{eq:two_u_relation}. Such zeros must exist due to $C > 0$, which can be justified using Lemma \ref{lem:ChernNumberZeros}.
However, the shape of the respective zeros are described by $\omega_\mu^{(i)}$, which should therefore be proportional to each other.
The multi-band nature of \eqref{eq:two_u_relation} makes it difficult to immediately conclude this, however, so we pass to ``Slater wavefunctions"
\begin{equation}
   \sU_{\sk^{(i)}}^{(i)}(\br_1, \ldots \br_N) = \varepsilon^{a_1, \ldots a_N} \su^{(i)}_{\sk a_1}(\br_1) \ldots \su^{(i)}_{\sk a_N}(\br_N),
\end{equation}
where $N$ is the number of Bloch bands (that comprise a single vortexable ``band"). Using \eqref{eq:two_u_relation} we obtain
\begin{equation}
\begin{aligned}
   & \sU_{\sk_1}^{(1)}(\br_1, \ldots \br_N) \\
   & = \det (\tilde{S}_\bk) e^{-i \bk \cdot \sum_{a=1}^N (\br'_1(\br_a) - \br'_2(\br_a))} 
   \sU_{\sk_2}^{(2)}(\br_1, \ldots \br_N).
   \end{aligned}
   \label{eq:slater_relation}
\end{equation}

Since $C > 0$, Lemma \ref{lem:ChernNumberZeros} implies that both $\sU^{(1)}_{\sk_1}$ and $\sU^{(1)}_{\sk_2}$ have exactly $C$ zeros in the Brilouin zone (we will only focus on one). In fact, due to the proportionality Eq. \eqref{eq:slater_relation}, the set of zeros and their multiplicities is the same for both wavefunctions. These zeros are in general dependent on $\{\br_i \}$; the real space coordinates should be understood as fixed in the discussion below. Let us choose one of the zeros $\bk_0$, which has the corresponding complex coordinates $\sk^{(i)}_0 = \sk^{(i)}(\v{k}_0)$.  As $\bk \to \v{k}_0$, we have
\begin{equation}
    \sU_{\sk^{(i)}}(\v{r}_1,\dots,\v{r}_N) 
    \propto (\sk^{(i)} - \sk^{(i)}_0)^{p} 
    = \left( 
    \epsilon^{\mu\nu} \omega_{\mu}^{(i)} [k_\nu - (k_0)_\nu]
    \right)^p.
\end{equation}
Combining this with Eq. \eqref{eq:slater_relation} and taking the limit $\v{k} \to \v{k}_0$, we must have

\begin{equation}
    \frac{\varepsilon^{\mu \nu}\omega^{(1)}_\mu \delta k_\nu}{\varepsilon^{\mu \nu}\omega^{(2)}_\mu \delta k_\nu} = \text{constant}, 
    \label{eq:omega_proportional}
\end{equation}
for all $\delta k_\nu = k_\nu - k_{0, \nu}$. This is only possible if $\omega^{(1)}_\mu \propto \omega^{(2)}_\mu$, which follows from  successively setting $\delta \bk = (1,0)$ and $\delta \bk = (0,1)$ in \eqref{eq:omega_proportional}.

We have nowhere set the scale of $\sz_i$ and $\omega^{(i)}_\mu$, so let us scale the vortex functions $\sz_i$ by a suitable complex number such that $\omega^{(1)}_\mu = \omega^{(2)}_\mu = \omega_\mu$ and $\sk_1 = \sk_2 = \sk$. The relationship \eqref{eq:two_u_relation} now takes the form
\begin{equation}
    \ket{\su^{(1)}_{\sk}} = e^{-i \bk \cdot (\br_1'- \br_2')} \tilde{S}_{\bk} \ket{\su^{(2)}_{\sk}}.
    \label{eq:two_u_relation_reduced}
\end{equation}

We act with $\ov{\partial}_\sk$:
\begin{equation}
\begin{aligned}
    0 = i\ov{\partial}_\sk \ket{\su^{(1)}_{\sk}} & = \omega_\mu [r_1^{\prime \mu} - r_2^{\prime \mu} - (\partial^{k_\mu} \tilde{S}_\bk)\tilde{S}_\bk^{-1} \ket{\su^{(1)}_{\sk}} \\
    & = (\sz_1 - \sz_2 - T_\bk) \ket{\su^{(1)}_{\sk}},
    \end{aligned}
\end{equation}
where we set $T_\bk = (\ov{\partial}_\sk \tilde{S}_\bk)\tilde{S}_\bk^{-1}$.

We wish to show that $T_\bk$ must be diagonalizable. Suppose not, and consider a sufficiently small region $B$ of $\bk$-space such that the eigenvalues of $T_\bk$, $\lambda_i(\bk)$, remain fixed in number and do not have crossings. Then we have a Jordan decomposition $T_\bk = V_\bk M_\bk V_\bk^{-1}$ where
\begin{equation}
M = \diag \begin{pmatrix}
    M_{\lambda_1}, M_{\lambda_2},\dots
\end{pmatrix}
\text{where }
    M_{\lambda} = \begin{pmatrix} 
    \lambda &  0 &  0    & \cdots   \\
    1       & \lambda & 0& \cdots  \\
    0    &     1     &\lambda & \cdots   \\
    \vdots &        &   & \ddots & 
    \end{pmatrix}.
    \label{eq:Jordan_decomposition}
\end{equation}
Let us focus on a single Jordan block of size $D \times D$ associated to an eigenvalue $\lambda(\bk)$. We assume and soon contradict $D>1$. Put $\ket{\tilde{u}^{(1)}_\bk} = V_\bk \ket{\su^{(1)}_\bk}$. Then in this block we have
\begin{equation}
    0 = \big[\sz_1(\br) - \sz_2(\br)\big] \tilde{u}^{(1)}_{\bk,i}(\br) + \sum_j M_{ij}(\bk) \tilde{u}^{(1)}_{\bk,j}(\br)
    \label{eq:Jordan_vortex_equality}
\end{equation}
for $1 \le i \le D$. For $i=1$, using the Jordan form \eqref{eq:Jordan_decomposition} we have
\begin{equation}
    0 = \left[\sz_1(\br) - \sz_2(\br) -\lambda(\bk)\right]
    \tilde{u}^{(1)}_{\bk,1}(\br).
    \label{eq:first_vortex_equality}
\end{equation}
such that whenever $\tilde{u}^{(1)}_{\bk,1}(\br) \neq 0$, for $\bk$ in our restricted region $B$,  we have $\sz_1(\br) - \sz_2(\br) = \lambda(\bk) = \lambda$. 

The equation for $i=2$ yields
\begin{equation}
    0 = \left[\sz_1(\br) - \sz_2(\br) -\lambda(\bk)\right]
    \tilde{u}^{(1)}_{\bk,2}(\br) + \tilde{u}^{(1)}_{\bk,1}(\br)
    \label{eq:jordan_contradiction}
\end{equation}
The result \eqref{eq:jordan_contradiction}, together with \eqref{eq:first_vortex_equality}, implies that $D=1$ such that $M$ is diagonal: indeed, choose some $\br = \br_0$ for which $\tilde{u}^{(1)}_{\bk_0,1}(\br_0) \neq 0$. Then using \eqref{eq:first_vortex_equality} we may set $\sz_1(\br_0) - \sz_2(\br_0) -\lambda(\bk) =0$ in \eqref{eq:jordan_contradiction} which contradicts our assumption that $M$ has a Jordan block of size larger than $1$.

We may then take $M = \diag(\lambda_1(\bk), \ldots, \lambda_N(\bk))$. For any $\br$ for which the density $\rho(\br) \neq 0$, which almost all $\br$ by assumption, there is a neighborhood of $\br$ (by continuity) such that $u^{(1)}_{\bk i}(\br) \neq 0$. Using \eqref{eq:Jordan_vortex_equality} for this $i$ we conclude $\sz_1(\br) - \sz_2(\br) = \lambda_i(\bk)$, or alternatively $\bnabla_\br(\sz_1(\br) - \sz_2(\br)) = 0 $ for almost all $\br$. Continuous functions cannot differ only on measure zero sets of $\mathbb{R}^2$ so we must have 
$\bnabla_\br(\sz_1(\br) - \sz_2(\br)) = 0 $ for all $\br$ such that $\sz_1(\br) - \sz_2(\br)$ differ only by a constant $\lambda$ as claimed.
\qed
}

\bigskip

We now conclude by restricting $\omega_\mu$ in systems with discrete $n>2$ fold rotation symmetry.

\prop{Consider a vortexable band with discrete translation symmetry $\sz(\br + \ba) = \sz(\br) + \omega_{\mu} a^\mu$, a finite continuous and generically nonzero electron density, and $C_{n>2}$ fold rotation symmetry. Then we can scale the vortex function such that $\sz(\br + \ba) = \sz(\br) + a_x + i\sign(\Omega) a_y$
\label{prop:rotation_symmetry_constraints}
}
\proof{
Without loss of generality we measure $\br$ from a center of rotation such that $\br \to C_{n} \br$ is the vector-action of rotations. From $[\P, C_n] = 0$, we have that $\sz(C_{n} \br)$ is a vortex function if $\sz(\br)$ is, such that $\sz(C_n \br) \propto \sz(\br)$ by uniqueness (Prop \ref{prop:uniqueness}). 
Such one dimensional representations of $C_n$ are labeled by an angular momentum $m$, modulo $n$, that counts the minimum number of times $\sz$ winds around $0$ upon a $2\pi$ rotation of $\br$. Since $\sz$ is injective as a consequence of \eqref{eq:non-degeneracy_condition}, the angular momentum must be $\pm1$.
We therefore conclude
\begin{equation}
    \sz(C_{n} \br) = e^{\pm 2\pi i/n} \sz(\br).
    \label{eq:rotation_vortex_representation}
\end{equation}
We will soon see that the $\pm$ sign corresponds to chirality $\sign(\Omega)$. (This is reasonable because it flips under $\sz \to \ov{\sz}$ and is positive for $\sz = x+iy$.)

Let us add a constant to $\sz$ such that $\sz(\v{r} = 0) = 0$. Then we have, using Prop. \ref{prop:general_translation},
\begin{equation}
    \sz(\ba) = \omega_\mu \ba^\mu.
\end{equation}
Combined with \eqref{eq:rotation_vortex_representation} we have
\begin{equation}
    \omega_\mu (C_{n} \ba)^\mu = e^{\pm 2\pi i/n} \omega_\mu \ba^\mu
    \label{eq:omega_map_rep}
\end{equation}
We claim that \eqref{eq:omega_map_rep} implies
\begin{equation}
\omega_\mu \ba^\mu = \gamma(a_x \pm i a_y)
\label{eq:omega_map_form}
\end{equation}
for some constant $\gamma$.
  A geometric picture is helpful here. Consider first the $+$ sign in \eqref{eq:omega_map_rep}. The complex vector $\omega_\mu$ can be understood as a linear map $\mathbb{R}^2 \to \mathbb{C} \simeq \mathbb{R}^2$ that commutes with rotation symmetry (counter-clockwise in both the domain and codomain). If it is not equal to a constant $\gamma$ times the identity, then it has two distinct eigenvectors must be each rotationally invariant which is impossible. For the $(-)$ rep, we may use the same argument above after applying an orientation reversal on one of the domain or codomain. 
  
  We now identify the $\pm$ sign in \eqref{eq:omega_map_form} with $\sign(\Omega)$. This follows from $\sign(-i \varepsilon^{\mu \nu} \ov{\omega_\mu} \omega_\nu) = \pm$ for $\omega = \begin{pmatrix}
      1 & \pm i
  \end{pmatrix}^T$ and the usage of Lemma \ref{lem:vortex_stokes}. Then taking $\sz \to \sz/\gamma$ completes the proof.
  \qed.
}

\bibliography{references_fixed}

\end{document}